\documentclass[journal]{IEEEtran}
\usepackage{graphicx}
\usepackage{amsmath}
\usepackage{algorithm}
\usepackage{hyperref}
\usepackage{booktabs} 
\usepackage{tabularx} 

\usepackage{booktabs} 
\usepackage{tabularx} 
\usepackage{caption}  
\usepackage{mathtools}
\usepackage{amssymb}
\usepackage{amsmath}

\usepackage{makecell}
\usepackage{textgreek}
\usepackage{comment}
\usepackage{bm}
\usepackage{booktabs}
\usepackage{color}
\usepackage{algorithmicx}
\usepackage{algpseudocode}
\usepackage{subcaption}
\usepackage[normalem]{ulem}
\usepackage{multirow}
\usepackage{colortbl}
\usepackage{array}
\usepackage{tabularx} 
\usepackage{makecell} 
\definecolor{lightblue}{rgb}{0.678, 0.847, 0.902} 

\algnewcommand\Input{\item[\textbf{Input:}]}  
\algnewcommand\Output{\item[\textbf{Output:}]}  
\usepackage{graphicx} 
\begin{document}

\title{Upward Spatial Coverage Recovery via Movable Antenna in Low-Altitude Communications}
 
\author{ Kan Yu,~\IEEEmembership{Member,~IEEE}, Kaixuan Li, Yujia Zhao, Dingyou Ma,~\IEEEmembership{Member,~IEEE}, Qixun Zhang,~\IEEEmembership{Member,~IEEE}, and Zhiyong Feng,~\IEEEmembership{Senior Member,~IEEE}
\thanks{This work is supported by the National Natural Science Foundation of China with Grants 62321001, 62301076, and 62401077, Fundamental Research Funds for the Central Universities with Grant  24820232023YQTD01,  Beijing Municipal Natural Science Foundation with Grant L232003, and National Key Research and Development Program of China with Grant 2022YFB4300403.
}

\thanks{K. Yu is with the Key Laboratory of Universal Wireless Communications, Ministry of Education, Beijing University of Posts and Telecommunications, Beijing, 100876, P.R. China. E-mail: kanyu1108@126.com;}
\thanks{K. Li is with the School of Computer Science, Qufu Normal University, Rizhao, P.R. China. E-mail: lkx0311@126.com;}
\thanks{Y. Zhao is with the Key Laboratory of Universal Wireless Communications, Ministry of Education, Beijing University of Posts and Telecommunications, Beijing, 100876, P.R. China. E-mail: yjzhao0318@126.com;}
\thanks{Y. Zhao is with the Key Laboratory of Universal Wireless Communications, Ministry of Education, Beijing University of Posts and Telecommunications, Beijing, 100876, P.R. China. E-mail: dingyouma@bupt.edu.cn;}
\thanks{Q. Zhang (\emph{the corresponding author}) is with the Key Laboratory of Universal
Wireless Communications, Ministry of Education, Beijing University of Posts and Telecommunications, Beijing, 100876, P.R. China. E-mail: zhangqixun@bupt.edu.cn;}
\thanks{Z. Feng (\emph{the corresponding author}) is with the Key Laboratory of Universal Wireless Communications, Ministry of Education, Beijing University of Posts and Telecommunications, Beijing, 100876, P.R. China. E-mail: fengzy@bupt.edu.cn.}

}

\markboth{IEEE Transactions on Vehicular Technology,~Vol.~, No.~, 2026}%
{Shell \Baogui Huang{\textit{et al.}}: Shortest Link Scheduling Under SINR}
\maketitle
\begin{abstract}
The rapid proliferation of unmanned aerial vehicle (UAV) applications imposes stringent requirements on continuous and reliable communication coverage in low-altitude airspace. Conventional cellular systems built upon fixed-position antennas (FPAs) are inherently constrained by static array geometries and limited mechanical degrees of freedom, which severely restrict their ability to adapt to highly dynamic three-dimensional (3D) propagation environments.
Movable antenna (MA) technology has recently emerged as a promising paradigm to overcome these limitations by actively reconfiguring electromagnetic radiation characteristics through controllable antenna positioning and array orientation, thereby enabling flexible spatial coverage adaptation. To systematically quantify the airspace coverage capability of MA-enabled systems, this paper formulates a spatial coverage maximization problem over a discretized 3D voxel space. For each voxel, the received signal-to-noise ratio (SNR) is maximized via joint optimization of the MA's 3D positions and beamforming matrices.
To efficiently solve the resulting non-convex problem, a hybrid particle swarm optimization and simulated annealing framework is developed to search for high-quality antenna configurations. Simulation results demonstrate that the proposed MA design framework substantially outperforms conventional FPA-based schemes in terms of spatial coverage, achieving coverage rates of 26.8\% and 29.65\%  for airspace below 300m and 600m, respectively. Moreover, further coverage enhancement can be attained by incorporating mechanical tilt adjustment, highlighting the strong potential of MA technology for reliable low-altitude communication coverage.

\end{abstract}
\begin{IEEEkeywords}
Movable antenna; Upward spatial coverage; Fixed-position antenna; Position and beamforming
\end{IEEEkeywords}

\IEEEpeerreviewmaketitle

\section{Introduction}

The rapid proliferation of unmanned aerial vehicles (UAVs) has fundamentally reshaped the paradigm of wireless communications by extending service demands from the terrestrial plane to the low-altitude three-dimensional (3D) space. Emerging UAV-enabled applications, such as logistics delivery, urban surveillance, traffic monitoring, and geospatial sensing, impose stringent requirements on communication reliability, continuity, and spatial coverage \cite{coverage2,coverage1}. Unlike conventional ground users, UAVs operate across a wide altitude range and exhibit high mobility, rendering wireless links highly sensitive to coverage discontinuities, abrupt channel degradation, and interference fluctuations. In this context, ensuring continuous and reliable \emph{spatial communication coverage} in low-altitude airspace has become a critical prerequisite for the safe and efficient operation of UAV networks.

In practice, most UAV missions are conducted within the altitude range of [0, 300]m, while certain applications, such as wide-area monitoring and tourism services, may extend to altitudes of up to $600$m \cite{range1}. However, existing cellular infrastructure is predominantly designed to serve terrestrial users, with base station (BS) antennas featuring downward-oriented radiation patterns and limited vertical adaptability. As a result, the upward spatial domain above BSs often suffers from systematic coverage deficiencies, manifested as tower-top blind zones, inter-site coverage gaps, and fragmented high-altitude connectivity \cite{range1,Wenxu1}. These deficiencies severely hinder the provision of seamless communication services in low-altitude airspace, making \emph{upward spatial coverage recovery} a fundamental challenge for future aerial–terrestrial integrated networks.

\begin{figure}[!ht]
    \centering
    \begin{subfigure}{\linewidth}
        \centering
        \includegraphics[width=0.65\linewidth]{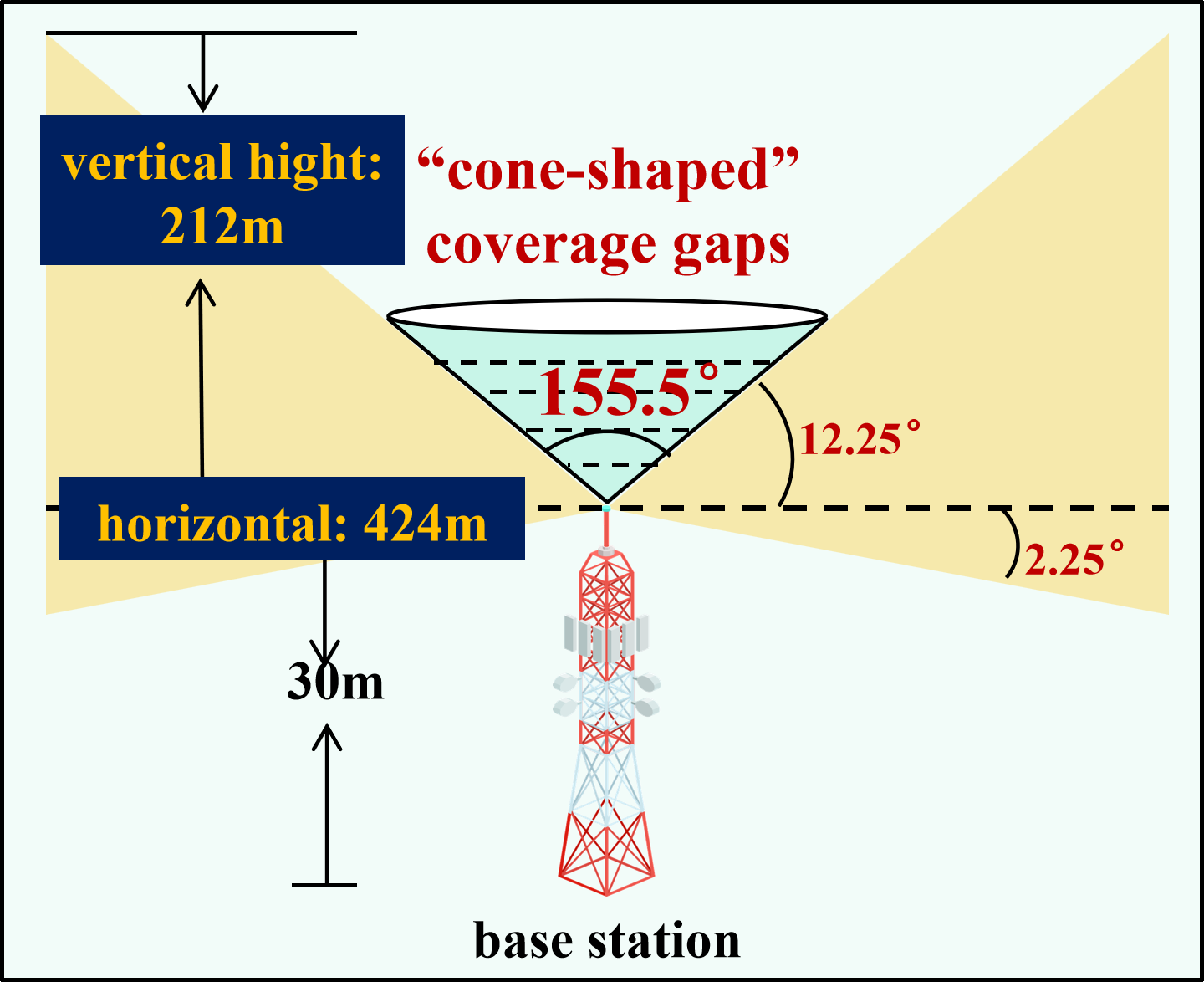}
        \caption{\small Coverage gaps for a single BS}
        \label{fig:gap1}
    \end{subfigure}
    \begin{subfigure}{\linewidth}
        \centering
        \includegraphics[width=0.7\linewidth]{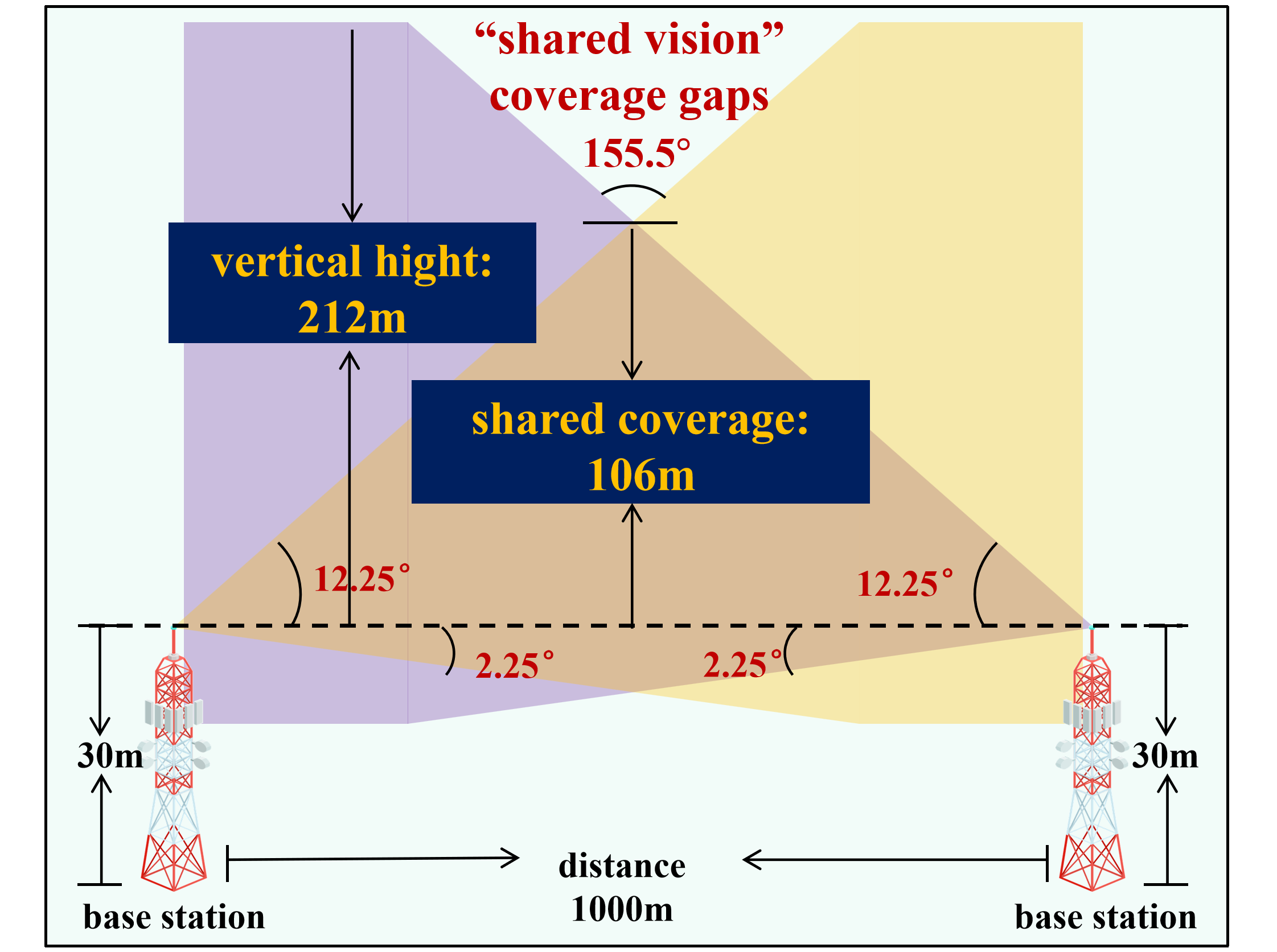}
        \caption{\small Coverage gaps for dual BSs}
        \label{fig:gap2}
    \end{subfigure}
    \caption{\small Diagram of BS coverage gaps (carrier frequency: $3.5$GHz, array configuration: $8 \times 8$ uniform planar array, beamwidth: $14.5^\circ$, Effective vertical scanning range: [$-10^\circ$, $+20^\circ$], mechanical downtilt: $-5^\circ$)
  }
    \label{fig:gpa}
\end{figure}

To effectively address low-altitude 3D coverage demands, existing solutions largely rely on fixed-position antenna (FPA) architectures and can be broadly categorized into three classes: mechanical antenna tilt deployment, vertical multi-beam transmission, and dual-carrier or multi-band co-deployment. Specifically, mechanical downtilt or uptilt strategies exploit antenna main lobes or side lobes to extend coverage toward aerial users; vertical beamforming techniques allocate distinct beams to different altitude layers; and dual-carrier schemes separate aerial and terrestrial services in the frequency domain. While these approaches partially alleviate coverage limitations, they remain inherently constrained by the static physical structure of FPAs. As illustrated in Fig.~\ref{fig:gpa}, the resulting spatial coverage is often characterized by cone-shaped blind zones above BSs and wedge-shaped gaps between neighboring cells. Moreover, spatial coverage based on side lobes is typically fragmented and power-limited, making it difficult to guarantee continuous connectivity at higher altitudes.

The fundamental limitation of FPA-based solutions lies not in algorithmic design, but in the \emph{lack of spatial degrees of freedom (DoF) at the hardware level}. Once deployed, the antenna geometry and radiation characteristics of FPAs are essentially fixed, allowing only limited adaptation through beamforming weights or frequency allocation. Consequently, existing methods can only passively compensate for coverage gaps, rather than actively reshaping the spatial radiation and channel conditions. This rigidity renders FPA-centric architectures ill-suited for highly dynamic low-altitude communication environments.

Recently, movable antenna (MA) technology has emerged as a promising paradigm to overcome these limitations by introducing controllable spatial DoF through antenna mobility \cite{Yu_Survey,Shao20256D}. Unlike traditional FPAs, MA systems enable dynamic adjustment of antenna positions and orientations in 3D space, thereby actively reconfiguring radiation patterns and wireless channels. By jointly optimizing antenna positioning and beamforming, MA systems can adapt to the electromagnetic environment, enhance link quality, suppress line-of-sight-dominated interference, and, critically, compensate for tower-top and inter-site coverage gaps. These capabilities suggest that MA technology \emph{has the potential to fundamentally reshape spatial coverage characteristics}, rather than merely extending coverage within the constraints of fixed antenna geometries.

Despite this potential, a systematic understanding of the coverage capabilities of MA-enabled systems remains largely unexplored. In particular, fundamental questions remain unanswered: \emph{\textbf{How does the spatial coverage capability of MA systems differ from that of conventional FPAs?}} and \emph{\textbf{What are the achievable coverage performance boundaries enabled by antenna mobility?}} Addressing these questions is essential for quantifying the true benefits of MA technology and guiding its practical deployment in low-altitude communication systems.

To bridge this gap, this paper presents the first systematic investigation of coverage capability differences between FPA- and MA-based architectures in low-altitude airspace. Using the received signal-to-noise ratio (SNR) as a unified performance metric, we formulate a spatial coverage maximization framework that explicitly captures the impact of 3D antenna positioning and beamforming. By discretizing the low-altitude region into volumetric elements, we optimize the MA positions and beamforming vectors to maximize coverage over the entire spatial domain, enabling a direct and quantitative comparison between FPA and MA solutions.
The main contributions of this paper are summarized as follows:
\begin{itemize}
    \item \textbf{Spatial coverage modeling:} We establish a low-altitude spatial coverage maximization framework that characterizes the coverage capability of a single MA array using an SNR-based voxel-level criterion, enabling fine-grained evaluation of upward coverage recovery.
    \item \textbf{Joint optimization framework:} The joint optimization of 3D MA positioning and beamforming is decomposed into tractable subproblems. A hybrid particle swarm optimization (PSO) and simulated annealing (SA) algorithm is developed for MA position optimization, while maximum ratio transmission (MRT) is employed for efficient beamforming design.
    \item \textbf{Performance comparison and insights:} Extensive simulations compare conventional FPA systems, MA systems with 3D positioning (i.e., 3DMA), and MA systems with additional mechanical tilt flexibility (i.e., 4DMA). The results demonstrate a clear coverage performance hierarchy, revealing the fundamental advantages of MA-enabled architectures in recovering upward spatial coverage.
\end{itemize}

The remainder of this paper is organized as follows. Section~\ref{sec:related_work} reviews existing spatial coverage enhancement techniques. Section~\ref{sec:network model} introduces the system model and formulates the spatial coverage maximization problem. Section~\ref{sec:opt} presents the proposed optimization framework and solution methodology. Section~\ref{sec:sim} provides comparative performance evaluations. Finally, Section~\ref{sec:conclusion} concludes the paper.

\begin{table*}[t] 
	\caption{Comparison Analysis of Different Coverage Schemes}
	\label{tab:FPA_works}
	\centering
	\arrayrulecolor{black}
	\arrayrulewidth=0.5pt
	\renewcommand{\arraystretch}{1.5} 
	
	\begin{tabular}{|>{\centering\arraybackslash}m{1.3cm}|  
			>{\centering\arraybackslash}m{1cm}|    
			>{\centering\arraybackslash}m{2.2cm}|   
			>{\centering\arraybackslash}m{1.0cm}|   
			>{\centering\arraybackslash}m{2cm}|     
			>{\raggedright\arraybackslash}m{3.5cm}| 
			>{\raggedright\arraybackslash}m{3.5cm}|}
		\hline  
		\textbf{Category} & 
		\textbf{Ref.} & 
		\textbf{Coverage scheme} & 
		\textbf{Coverage height} & 
		\textbf{Metrics} & 
		\textbf{Advantages} & 
		\textbf{Shortages} \\ 
		\hline
		Sidelobe & \cite{Cellular2019} & Array downtilt,\newline Sidelobe Coverage & $\leq$300m & SINR & 
		1) Low cost;\newline 2) Fast deployment;\newline 3) High resource reuse rate & 
		1) Sidelobe increases interference;\newline 2) Severe fragmentation;\newline 3) Numerous coverage gaps \\  
		\hline
		\multirow{2}{1.3cm}{\centering Antenna uptilt} & \cite{Network2019} & Array uptilt & $\leq$300m & 3D coverage & 
		\multirow{2}{3.5cm}{1) Low cost;\newline 2) Fast deployment;\newline 3) High resource reuse rate} & 
		\multirow{2}{3.5cm}{1) Rigid mechanical structure;\newline 2) Numerous coverage gaps;\newline 3) Weak signal at receiver} \\  
		\cline{2-5}
		& \cite{Maeng} & Downtilt regulation & 300m & SINR, outage probability &  &  \\ 
		\hline
		\multirow{2}{1.3cm}{\centering Multi-beam} & \cite{Research2025} & Vertical layering + beam design & $\leq$1000m & Signal/Capacity coverage, energy efficiency & 
		\multirow{2}{3.5cm}{1) High coverage strength and quality;\newline 2) Balances air-ground coverage;\newline 3) Flexible beam regulation} & 
		\multirow{2}{3.5cm}{1) Rigid mechanical structure;\newline 2) Existing coverage gaps;\newline 3) Limited antenna tilt angle} \\  
		\cline{2-5}
		& \cite{Telecommunication2022} & "1+X" + "7 horizontal beams" & None & Coverage range,\newline Signal quality&  &  \\ 
		\hline
		Single/dual carrier & \cite{Wei1,Wei2} & "3+4" + dual carrier & $\leq$300m & Coverage rate, RSRP, SINR & 
		1) Simultaneous air-ground coverage;\newline 2) Large capacity with dual carrier scheme;\newline 3) Inter-frequency scheme avoids interference & 
		1) Rigid mechanical structure;\newline 2) Existing blind spots;\newline 3) Challenging to balance capacity-interference-cost requirements \\  
		\hline
		MA-enabled  & Proposed in this paper& MA positions and up-tilt control & $\leq$600m & Coverage rate, SNR & 
		1) High flexibility with rich spatial DoFs;\newline 2) Strong channel reuse capability;\newline 3) Adjustable mechanical tilt;\newline 4) Low energy consumption;\newline 5) Interference avoidance at the receiver & 1) Mechanical delay exists \\  
		\hline
	\end{tabular}
	\\[0.6cm]
	\raggedright
	\textbf{Note:} Ref.: Reference; RSRP: Reference signal received power; SINR: Signal to interference plus noise ratio.
\end{table*}

\section{Related Works}\label{sec:related_work}
The antennas equipped with conventional cellular BS are predominantly designed with downward-oriented radiation patterns to guarantee terrestrial coverage, while aerial communication has long been treated as a secondary or incidental service. As a consequence, the upward spatial domain often suffers from systematic coverage deficiencies, which we refer to as the \emph{upward coverage gap}. In this section, we review representative studies on cellular-based spatial coverage enhancement, with a focus on FPA solutions and their intrinsic limitations. These limitations further motivate the adoption of MA technology for upward coverage recovery, which constitutes the core focus of this work.

\subsection{Elevation-Aware Coverage Enhancement with FPA}
Most existing studies on cellular-enabled spatial coverage enhancement rely on FPAs and can be broadly classified into three categories: \emph{mechanical downtilt (or uptilt) control}, \emph{vertical multi-beam design}, and \emph{dual-carrier or frequency-domain} solutions. A qualitative comparison of representative FPA-based approaches is summarized in Table~\ref{tab:FPA_works}.

\subsubsection{Mechanical Downtilt and Uptilt Control}

To preserve terrestrial performance while partially supporting aerial users, early studies exploited the upper side lobes of downtilted base station antennas to provide connectivity to UAVs, owing to their low deployment cost and implementation simplicity. Xu \emph{et al.} incorporated realistic antenna radiation patterns into the system model and demonstrated that side-lobe-based aerial coverage inevitably introduces additional interference to UAV communications \cite{Cellular2019}.

However, side-lobe-assisted schemes inherently suffer from fragmented spatial coverage and insufficient link budgets due to low side-lobe gain, which limits their ability to support altitude-differentiated services. A more direct way is to mechanically up-tilt the FPA array to steer the main lobe toward the airspace. For instance, in~\cite{Dedicating2023}, Chen \emph{et al.} investigated a hybrid cellular deployment consisting of both up-tilted and down-tilted BSs, and showed that an appropriate proportion between the two can balance aerial and terrestrial coverage. Similarly, in~\cite{Network2019}, Lyu \emph{et al.} analyzed spatial coverage performance under negative downtilt configurations using a non-uniform antenna radiation model and derived analytical expressions for uplink and downlink coverage probabilities within the 40--300m altitude range. In \cite{Maeng}, Maeng \emph{et al.} further optimized antenna uptilt angles to maximize coverage probability within predefined UAV corridor regions.

Despite their effectiveness, mechanical tilt-based frameworks are constrained by practical deployment considerations. The feasible tilt range is typically limited, and once deployed, mechanically tilted antennas cannot be adaptively reconfigured, which restricts their applicability in highly dynamic low-altitude communication scenarios.

\subsubsection{Vertical Multi-Beam Design}

To overcome the limited flexibility of single-beam radiation patterns in the vertical domain, several works have proposed vertical multi-beam designs based on synchronization signal block (SSB) layering. For instance, in \cite{Research2025}, Yu \emph{et al.} optimized vertical SSB beam arrangements to mitigate coverage fragmentation in low-altitude airspace and introduced a multi-layer frequency reuse mechanism to alleviate co-channel interference. By leveraging large-scale FPA arrays, narrow vertical beams can be sequentially transmitted to cover different altitude layers.

In industrial practice, ZTE Corporation proposed the ``$1+N$'' vertical coverage mode, where one wide beam serves terrestrial users and $N$ beams target aerial users~\cite{ZTE2020}. While in~\cite{Telecommunication2022}, Li \emph{et al.} compared ``$1+X$'' configurations with conventional multi-beam schemes and demonstrated improvements in interference control and energy efficiency. Similarly, the Shanghai Vehicle Network Association suggested a ``$4+3$'' beam configuration to support aerial–terrestrial coverage~\cite{Technical2025}.

Although vertical multi-beam designs enhance altitude-dependent coverage, their effectiveness is primarily limited to relatively low altitudes (e.g., below 120m). In dense urban environments, non-line-of-sight (NLoS) conditions and cell-edge effects may still result in coverage discontinuities. Moreover, due to the limited electrical and mechanical tilt ranges of FPAs, vertical multi-beam schemes face an inherent tradeoff between coverage continuity and interference suppression.

\subsubsection{Dual-Carrier and Frequency-Domain Solutions}

Another line of research focuses on dual-carrier or multi-band deployment to separate aerial and terrestrial services. For instance, in~\cite{Lin2025}, Lin \emph{et al.} considered upgrading selected BSs to support aerial communications along predefined flight paths using a semi-supervised graph convolutional network. From an industrial perspective, China Mobile~\cite{ChinaM2021} recommended using different frequency bands for distinct altitude ranges, while China Telecom primarily relies on the 3.5GHz band for aerial–terrestrial integrated networking. Additional dual-carrier schemes, including frequency-division and co-frequency deployments, were discussed in~\cite{China2024}.

While frequency-division schemes effectively isolate aerial and terrestrial interference, they require additional spectrum resources and incur high deployment costs. Co-frequency schemes, on the other hand, suffer from severe interference and may degrade terrestrial network capacity. In \cite{Wei1,Wei2}, hybrid approaches combining frequency-division and co-frequency transmission have shown performance gains, yet they still rely on fixed antenna geometries and cannot fundamentally resolve upward coverage deficiencies.

\emph{\textbf{Summary of FPA-based spatial coverage enhancement}:}
Overall, existing FPA-based solutions enhance upward coverage primarily through beam shaping or spectrum partitioning, while the spatial DoFs of the FPA array itself remain fixed. As a result, these approaches struggle to simultaneously achieve continuous altitude-adaptive coverage, effective interference control, and flexible service support in highly dynamic low-altitude environments.

\subsection{Motivation for MA-Enabled Spatial Coverage Recovery}

The above studies reveal a common limitation of FPA-based approaches: coverage enhancement is achieved in a passive and indirect manner, constrained by static antenna geometries and limited tilt ranges. Such mechanisms are insufficient for addressing coverage holes, fragmented connectivity, and strong line-of-sight-dominated interference in low-altitude airspace.

To overcome these limitations, MA technology has recently emerged as a promising paradigm for flexible spatial coverage reconfiguration. For instance, in \cite{Flexible2024}, Wang \emph{et al.} jointly optimized MA positions and horizontal beamforming weights to maximize the minimum beam gain over designated spatial regions. In \cite{Gao_2026}, Gao \emph{et al.} further investigated MA-enabled communications assisted by reconfigurable intelligent surfaces, highlighting the tradeoff between MA flexibility, system complexity, and performance. Focusing on aerial scenarios, in~\cite{6DRen}, Ren \emph{et al.} optimized 6DMA positions, beamforming, and BS association to enhance the quality of SINR, demonstrating the effectiveness of MA techniques in mitigating LoS-dominated interference.

Collectively, these works validate the potential of MA technology in coverage extension and flexible service provisioning. However, existing studies primarily focus on planar or region-based coverage enhancement, while MA-enabled elevation-aware spatial coverage enhancement from a continuous low-altitude volumetric perspective remains largely unexplored. This gap directly motivates the present work.

\textbf{Notations}: In this paper, $ (\cdot)^{\mathbf{H}}$ and  $ (\cdot)^{\rm \mathbf{T}}$ denote the conjugate transpose and transpose operations, respectively. $\left\|\cdot \right\|_2$ denotes the Euclidean norm. $\odot$ denotes the Hadamard products. ${\mathbb{R}^{a \times b}}$ and ${\mathbb{C}}^{a \times b}$ are the set of $a \times b$ dimensional complex
and real matrices,

\begin{figure}
\centering
\includegraphics[width=3in]{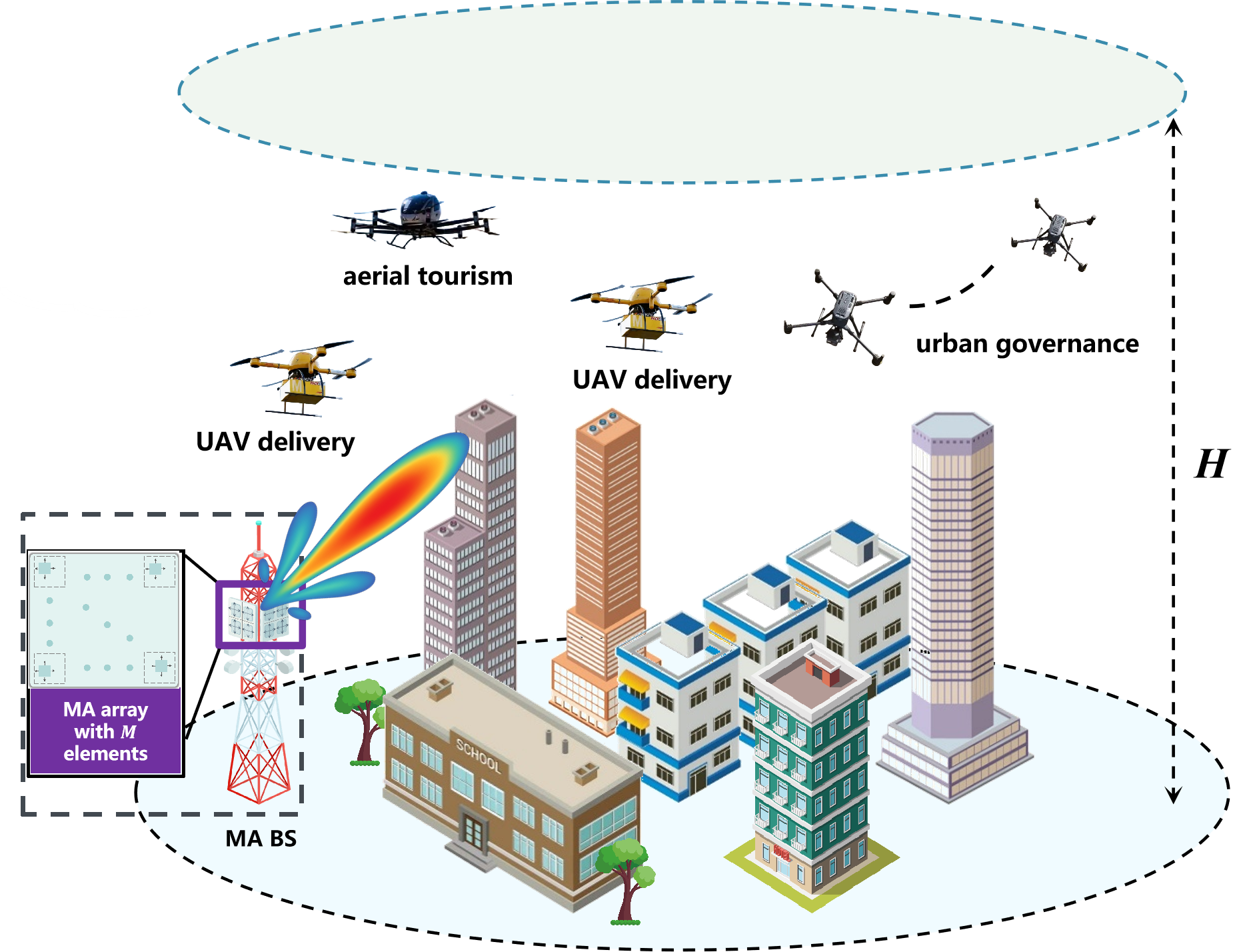}
\caption{\small Network model}
\label{fig:scenario1}
\end{figure}

\section{Network Model and Preliminaries}\label{sec:network model}

We consider a BS–UAV communication scenario, as shown in Fig. \ref{fig:scenario1}, where the BS is equipped with a MA array comprising 
$M$ elements. The MA elements form a planar transmit array with a mechanical downtilt angle $\beta$, and each antenna element is allowed to move within a prescribed 3D spatial region. Let $\bf{q}_m^{\rm LCS}$ and $\bf{q}_m^{\rm GCS}$ denote the position vectors of the $m$-th MA element in the local coordinate system (LCS) and the global coordinate system (GCS), respectively.
The UAV is assumed to operate within a predefined cubic airspace and is equipped with a single FPA for signal reception. 

\begin{figure}
\centering
\includegraphics[width=3in]{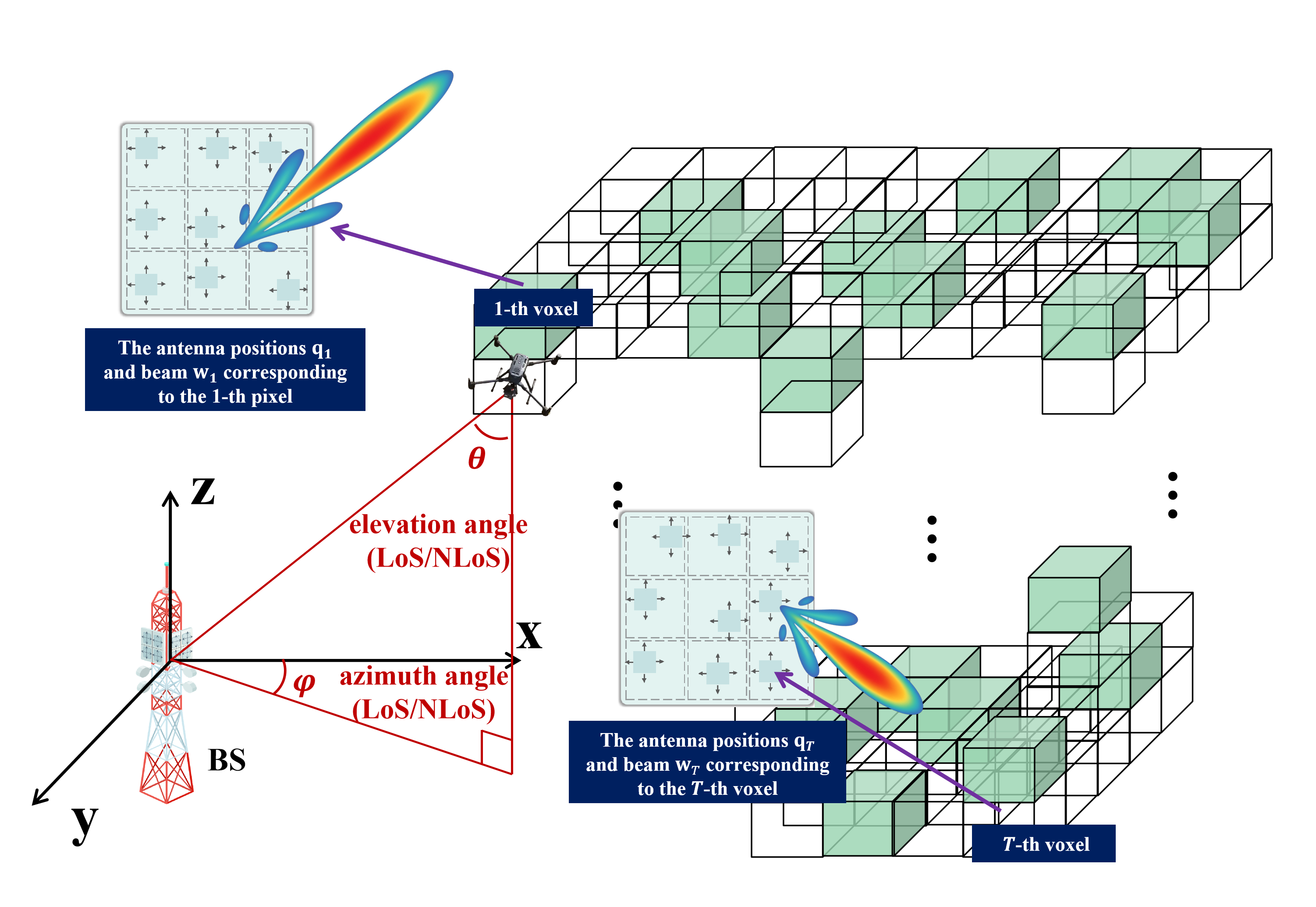}
\caption{\small Spatial gridding model}
\label{fig:scenario2}
\end{figure}

\subsection{Spatial Coverage Gridding}\label{sub: airspace gridding}
As shown in Fig. \ref{fig:scenario2}, the continuous 3D airspace is discretized into regular grid cells, resulting in a digital representation of the airspace. Considering a 3D coverage region with volume $V$, which is divided into $\rm{T}=\{1,\ldots,t,\ldots,T\}$ equal-sized voxels. The volume of each voxel is denoted by $\Omega=V/T$. It is assumed that the UAV is positioned at the center of each voxel, with the UAV's position in the $t$-th voxel represented by $\mathbf{p}_t$. To evaluate the coverage performance of the BS with MA array, the SNR is used to quantify the signal strength at each voxel\footnote{The reason for considering only the SNR and not accounting for interference in this scenario is due to the assumption of a relatively low interference environment, where the main focus is on the direct signal strength from the BS to the UAV. In practical low-altitude communication scenarios, especially when the BS and UAV are in close proximity and the system design prioritizes line-of-sight (LoS) communication, interference from other sources can be negligible compared to the primary signal.
Moreover, the SNR serves as a reasonable approximation of the system's overall communication performance, particularly for coverage analysis. In this context, the goal is to model the basic feasibility of communication over a given region, and interference, though important in more complex or dense networks, may not significantly alter the fundamental coverage assessment in the simplified model being considered here.}. A voxel is considered \emph{covered} if the SNR at the UAV in the $t$-th voxel, namely ${\rm UAV}_t$, exceeds a predefined threshold $\epsilon$; otherwise, the voxel is classified as \emph{uncovered}.

\subsection{Communication model}
\begin{figure}
\centering
\includegraphics[width=3in]{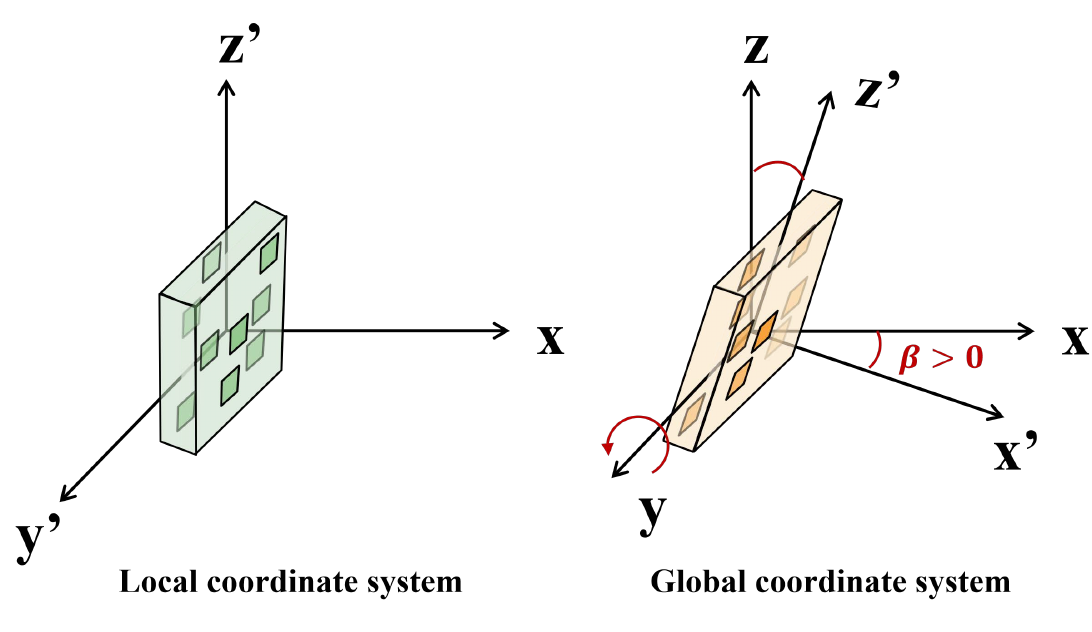}
\caption{\small The diagram of mechanical downtilt}
\label{fig:tilt}
\end{figure}

In fact, enhancing spatial coverage height or filling coverage blind spots using the MA is closely tied to the tilt angle of the MA array. As illustrated in Fig. \ref{fig:tilt}, the MA array at the BS can rotate along the $y$-axis to improve coverage. However, due to the mechanical constraints of the MA, when the array tilts downward, the mechanical downtilt angle 
$\beta$ becomes positive ($\beta>0$); conversely, when the array tilts upward, $\beta$ is negative ($\beta<0$).
The corresponding rotation matrix that governs this tilting operation can be represented as \cite{6DRen,Shao20256D}
\begin{equation}
    R=\left [ {\begin{matrix}\cos{\beta}&0&\sin{\beta}\\0&1&0\\$-$\sin{\beta}&0&\cos{\beta}\end{matrix}}\right ].
\end{equation}
Therefore, the position of the $m$-th antenna in the GCS is  given by $\mathbf{q}^{\rm GCS}_m =R \mathbf{q}^{\rm LCS}_m$. 

The communication path from the BS to the UAV located in the $t$-th voxel denoted as ${\rm UAV}_t$, consists of LoS link and non-LoS (NLoS) links. Specifically, the LoS path is represented by $L_{\text{LoS}} = 1$, while the NLoS paths are represented by $L_{\text{NLoS}}$. The azimuth angle and elevation angle of the $l$-th path are represented as $\varphi_l \in [-\frac{\pi}{2}, \frac{\pi}{2}]$ and  $\theta_l \in [0, \frac{\pi}{2}]$, respectively.  The corresponding pointing vector for the $(\varphi_l, \theta_l)$ pair is denoted by $f_l = [\sin \theta_l \cos \varphi_l, \sin \theta_l \sin \varphi_l, \cos \theta_l]^T$. For both LoS and NLoS links, the steering vector of the $m$-th antenna is expressed as \cite{Yujia_TCOM}
\begin{equation}
    {a_{L,m}}={{[{e^{i\frac{2π}{λ}{f_1^{\mathrm{T}}}{{\mathrm{q}}_m^{\mathrm{GCS}}}}},...,{e^{i\frac{2π}{λ}{f_L^{\mathrm{T}}}{{\mathrm{q}}_m^{\mathrm{GCS}}}}}]}^{\mathbf{\rm T}}}, 
\end{equation}
where $L \in \{ L_{\rm LoS}, L_{\rm NLoS}\}$. Then, the transmit field response matrix can be represented as 
\begin{equation}
    A_L={{[a_{L,1},...,a_{L,M}]}^{\mathbf{T}}} \in{{\mathbb{\mathbb{C}}}^{L\times M}}.
\end{equation}
The UAV is equipped with a fixed antenna, then the receiving steering vector is denoted by
\begin{equation}
   F_L={{[1,...,1]}^{\mathbf{T}}}\in{ {\mathbb{R}}^{L\times1}}.
\end{equation}

For the NLoS channel matrix from the BS to the UAV, we define $\Sigma_{\text{NLoS}} = \text{diag}(\sigma_1, \dots, \sigma_{L_{\rm NLoS}}) \in \mathbb{C}^{L_{\text{NLoS}} \times L_{\text{NLoS}}}$, where the diagonal elements are mutually independent and follow the same distribution, $ \mathcal{C}\mathcal{N}(0, \beta_0 d^{-\alpha L_{\text{NLoS}} ^{- 1}})$. For the LoS link, the channel matrix is given by $\Sigma_{\text{LoS}} = \beta_0 d^{-\alpha}$, where $\beta_0 = \left(\frac{\lambda}{4\pi}\right)^2$ represents the reference path loss at a standardized distance of 1 meter, and $d$ is the distance from the BS to the UAV located in the voxel.

\begin{figure*}[t] 
\centering
\begin{equation}  \label{eq:G}
G(\varphi, \theta, \beta) = 
\begin{cases} 
G_\rho \sin^\rho(\theta - \beta) \cos^\rho(\varphi), & \theta \in \left[0, \frac{\pi}{2}\right], \varphi \in \left[-\frac{\pi}{2}, \frac{\pi}{2}\right] \\
0, & \text{otherwise}
\end{cases} 
\end{equation}
\hrule 
\end{figure*}

Considering the impact of the directional antenna orientation on signal propagation, coverage, and communication performance, a cosine radiation pattern is adopted for modeling. Specifically, given the azimuth angle $\varphi$, elevation angle $\theta$, and downtilt $\beta$, the radiated power is expressed as $G(\varphi, \theta, \beta)$, as shown in Eq. \eqref{eq:G} \cite{G}. In this model, $G_\rho=2(1+\rho)$ represents the normalized power, $\rho$ is the directional pattern sharpening factor. Furthermore, the channel matrix for the NLoS links is given by $G_{\text{NLoS}} \in \mathbb{R}^{M \times L_{\text{NLoS}}}$. 
Therefore, the channel gain for NLoS links is denoted as \cite{GV2,GV1}
\begin{equation}
    H_{\text{NLoS}} = \sqrt{G_{\text{NLoS}} }\odot A_{\text{NLoS}}^T \Sigma_{\text{NLoS}}^T F_{\text{NLoS}} \in \mathbb{C}^{M \times 1}.
\end{equation}
Similarly, the channel gain for the LoS link is represented as 
\begin{equation}
H_{\text{LoS}} = \sqrt{ G_{\rm{LoS}}} \odot A_{\text{LoS}}^T \Sigma_{\rm{LoS}}^T F_{\rm{LoS}}.
\end{equation}
To sum up, the channel vector between the BS and the UAV in a Rician fading channel can be represented as \cite{Ricain1,Ricain2}
\begin{equation}
\bf{H}=\sqrt{\frac{\kappa}{\kappa+1}}{H_{\rm LoS}}+\sqrt{\frac{1}{\kappa+1}}H_{\rm NLoS },
\end{equation}
where the symbol $\kappa$ is the Rician factor. Therefore, the total received signal at the receiver is given by
\begin{equation}
  y = \mathbf{Hw} s+n,
\end{equation}
where $\bf{ w}$ is the beamforming vector at the transmitter, $s$ is the data stream transmitted by the base station, and $n$ is the noise signal, which follows a complex Gaussian distribution with zero mean and variance \( \sigma_n^2 \) \cite{Kaixuan_TMC}. The received SNR at the receiver is then given by
\begin{equation}
\text{SNR} = \frac{ {|\mathbf{H}^{\mathbf{\rm H}}\mathbf{w} |}^2 }{\sigma_n^2}.
\end{equation}
The reason that not accounting for interference has been given in Subsection \ref{sub: airspace gridding}. 

\subsection{Optimization Problem Formulation}

Motivated by the fundamental differences in spatial coverage behavior between FPAs and MAs in 3D airspace, we formulate a coverage evaluation problem to characterize the volumetric coverage capability of a single MA array. In this work, the mechanical downtilt of the antenna array is assumed to be fixed, and the corresponding coverability is evaluated under this configuration.

For a given UAV located at position $\mathbf{p}_t$ in the $t$-th voxel, the \emph{3D antenna positions} $\mathbf{q}_t = [\mathbf{q}^{\rm GCS}_{1},...,\mathbf{q}^{\rm GCS}_{M}]$ of the MA array and the \emph{transmit beamforming vector} $\mathbf{w}_t$ at the BS are jointly optimized. A location is deemed \emph{coverable} if there exists a feasible pair $(\mathbf{q}_t, \mathbf{w}_t)$ such that the resulting received SNR exceeds a prescribed threshold $\epsilon$; otherwise, the location is regarded as uncovered. This coverability condition is formally defined as
\begin{equation}
c_t =
\begin{cases}
1, & \gamma_t(\mathbf{p}_t, \mathbf{q}_t, \mathbf{w}_t) \ge \epsilon, \\
0, & \text{otherwise},
\end{cases}
\end{equation}
where $\gamma_t$ denotes the SNR received at the UAV located in the $t$-th voxel.
Based on the above definition, the volumetric coverability over the considered region is quantified as
\begin{equation}
\textbf{Pro}_{\rm vol} = \frac{\sum_{t=1}^{T} c_t}{V} \times \Omega,
\end{equation}
where $T$ is the total number of voxels, $V$ denotes the volume of the considered region, and $\Omega$ is the volume of each voxel.

Our objective is to traverse all $T$ voxels within the feasible region and optimize the MA positions $\mathbf{q}=\{\mathbf{q}_1,\ldots,\mathbf{q}_T\}$ together with the corresponding transmit beamforming vectors $\mathbf{W}=\{\mathbf{w}_1,\ldots,\mathbf{w}_T\}$, so as to maximize the overall volumetric coverability. This leads to the following optimization problem:
\begin{subequations}\label{eq:P1}
\begin{align}
\mathcal{P}_1:\quad 
& \max_{\mathbf{q}, \mathbf{W}} \ \textbf{Pro}_{\rm vol} \label{P1-obj} \\
\text{s.t.}\quad 
& \mathbf{q}_{t,m} \in \Psi_m, \quad m \in \{1,\ldots,M\}, \label{P1-1} \\
& \left\| \mathbf{q}_{t,m} - \mathbf{q}_{t,n} \right\|_2 \ge d_{\min}, 
 m,n \in \{1,\ldots,M\}, \ m \neq n, \label{P1-2} \\
& \left\| \mathbf{w}_t \right\|_2^2 \le P_{\max}, \label{P1-3}
\end{align}
\end{subequations}
where constraint~\eqref{P1-1} restricts each MA element to move within its prescribed feasible region $\Psi_m$, constraint~\eqref{P1-2} enforces a minimum inter-element distance to avoid physical collision, and constraint~\eqref{P1-3} imposes the transmit power constraint, with $P_{\max}$ denoting the maximum allowable transmit power.
It is worth noting that the objective function \eqref{P1-obj} in $\mathcal{P}_1$ is discontinuous due to the binary nature of $c_t$, and the optimization variables are strongly coupled through the SNR expression. As a result, problem $\mathcal{P}_1$ is highly non-convex and intractable to solve directly. 

\section{Joint Antenna Position and Beamforming Optimization for Volumetric Coverage}\label{sec:opt}
In this section, we develop an efficient algorithmic framework to solve the volumetric coverage maximization problem formulated in Eq. \eqref{eq:P1}. 
Due to the discrete nature of the coverage indicator and the strong coupling between the MA positions and transmit beamforming vectors, the original problem is highly non-convex and cannot be solved directly. 
To address this challenge, we first simplify the problem and then adopt a decomposition-based optimization strategy.

\subsection{Problem Simplification and Decomposition}
Since the total region volume $V$ and the voxel volume $\Omega$ are fixed, maximizing the volumetric coverability $\textbf{Pro}_{\rm vol}$ is equivalent to maximizing the number of coverable voxels, i.e., $\sum_{t=1}^{T} c_t$. As a result, the original coverage maximization problem can be decomposed into a set of voxel-wise feasibility checks.
For a given UAV located at $\mathbf{p}_t$, the corresponding voxel is considered covered if there exists a feasible pair $(\mathbf{q}_t, \mathbf{w}_t)$ such that the received SNR at ${\rm UAV}_t$ exceeds the prescribed threshold $\epsilon$. This formulation is justified by the fact that voxel coverability is determined solely by whether joint MA positioning and beamforming can drive the received SNR beyond the target threshold. Accordingly, for each voxel, the coverage decision can be obtained by solving the following SNR maximization problem:
\begin{subequations} \label{eq:P2}
\begin{align}
\mathcal{P}_2:\quad 
& \max_{\mathbf{q}_t,\mathbf{w}_t} \ \gamma_{t}(\mathbf{p}_t,\mathbf{q}_t,\mathbf{w}_t) \\
\text{s.t.}\quad 
& \mathbf{q}_{t,m} \in \Psi_m, \quad m \in \{1,\ldots,M\}, \\
& \left\| \mathbf{q}_{t,m} - \mathbf{q}_{t,n} \right\|_2 \ge d_{\min}, \quad m \neq n, \\
& \left\| \mathbf{w}_t \right\|_2^2 \le P_{\max}.
\end{align}
\end{subequations}

Problem $\mathcal{P}_2$ is highly non-convex due to the strong coupling between the MA positions $\mathbf{q}_t$ and the beamforming vector $\mathbf{w}_t$ in the SNR expression. Moreover, the objective function generally exhibits multiple local optima induced by the spatial geometry and antenna movement constraints. Therefore, directly solving $\mathcal{P}_2$ using conventional convex optimization techniques is intractable. To address this challenge, we decompose $\mathcal{P}_2$ into two subproblems: \emph{antenna position optimization} and \emph{beamforming optimization}, which are solved in an alternating manner.

\subsection{Antenna Position Optimization Subproblem}
For the SNR maximization problem associated with the $t$-th voxel, we first focus on optimizing the MA positions, while temporarily fixing the beamforming strategy. Due to the non-convex and geometry-dependent nature of the antenna position optimization problem, a PSO algorithm combined with SA is employed to search for high-quality solutions.

\subsubsection{PSO Algorithm}
It is utilized to iteratively update the MA positions so as to enhance the received SNR at ${\rm UAV}_t$. A swarm consisting of $K$ particles is initialized, where each particle represents a feasible realization of the MA array positions within the prescribed movement region.

Specifically, the initial positions and velocities of the swarm are given by $\{\mathbf{q}_k^{(0)}\}_{k=1}^{K}$ and $\{\mathbf{v}_k^{(0)}\}_{k=1}^{K}$, respectively, where
\begin{equation}
\mathbf{q}_k^{(0)} =
\begin{bmatrix}
\underbrace{x_{k,1}^{(0)},\ y_{k,1}^{(0)},\ z_{k,1}^{(0)}}_{\text{1-th MA}},\,
\ldots,\,
\underbrace{x_{k,M}^{(0)},\ y_{k,M}^{(0)}, z_{k,M}^{(0)}}_{\text{$M$-th MA}}
\end{bmatrix}^{\mathrm{T}}.
\end{equation}
where $(x_{k,m}^{(0)}, y_{k,m}^{(0)}, z_{k,m}^{(0)})$ denotes the 3D coordinates of the $m$-th MA element associated with the $k$-th particle, subject to the feasible region constraint in \eqref{P1-1}. The initial particle velocities are set to zero.

Let $\mathbf{q}_k^{(i)}$, $\mathbf{q}_{k,\mathrm{lbest}}$, and $\mathbf{q}_{k,\mathrm{gbest}}$ denote the current position, local best position, and global best position of the $k$-th particle at the $i$-th iteration, respectively. The velocity update rule is given by
\begin{equation}\label{eq:v_k}
\mathbf{v}_k^{(i+1)} = \omega^{(i)} \mathbf{v}_k^{(i)} 
+ c_1 s_1 \big( \mathbf{q}_{\mathrm{gbest}} - \mathbf{q}_k^{(i)} \big)
+ c_2 s_2 \big( \mathbf{q}_{k,\mathrm{lbest}} - \mathbf{q}_k^{(i)} \big),
\end{equation}
where $c_1$ and $c_2$ are acceleration coefficients, $s_1$ and $s_2$ are random vectors uniformly distributed in $[0,1]$, and $\omega^{(i)}$ denotes the inertia weight at the $i$-th iteration, which is updated as
\begin{equation}
\omega^{(i)} = \omega_{\max} - \frac{\omega_{\max} - \omega_{\min}}{I_{\max}}\cdot i,
\end{equation}
where $\omega_{\max}$ and $\omega_{\min}$ denote the maximum and minimum values of the inertia weight, respectively, and $I_{\rm max}$ represents the maximum number of iterations. 

To ensure the feasibility that antenna positions keep within the feasible region and avoid particles drifting outside the search space, boundary constraints are imposed on both particle velocities and positions \cite{PSO1,PSO2}
\begin{equation}\label{eq:v_cons}
\mathbf{v}_k^{(i+1)} = \max\!\left( \min(\mathbf{v}_k^{(i+1)}, \mathbf{v}_{\max}), -\mathbf{v}_{\min} \right),
\end{equation}
\begin{equation}\label{eq:q_k}
\mathbf{q}_k^{(i+1)} = \max\!\left( \min(\mathbf{q}_k^{(i)} + \mathbf{v}_k^{(i+1)}, \mathbf{q}_{\mathrm{ub}}), \mathbf{q}_{\mathrm{lb}} \right),
\end{equation}
where ${\mathbf{ v}}_{\max}$ and ${\mathbf{ v}}_{\min}$ are the maximum and minimum velocity constraints, respectively, while $\mathbf{q}_{\rm ub}$ and $\mathbf{q}_{\rm lb}$ represent the upper and lower bounds of the particle positions.

\subsubsection{SA Enhancement}

To mitigate the risk of PSO being trapped in local optima, the SA mechanism is incorporated to allow occasional acceptance of inferior solutions with a certain probability, thereby enhancing the global exploration capability. Specifically, the position update is accepted according to
\begin{equation}
\mathbf{q}_k^{(i+1)} =
\begin{cases}
\mathbf{q}_k^{(i+1)}, & \gamma_k^{(i+1)} > \gamma_k^{(i)} \ \text{or accepted w.p. } p_{\mathrm{SA}}^{(i)}, \\
\mathbf{q}_k^{(i)}, & \text{otherwise},
\end{cases}
\end{equation}
where the acceptance probability is given by
\begin{equation}\label{eq:p_SA}
p_{\mathrm{SA}}^{(i)} = \exp\!\left( \frac{\gamma_k^{(i+1)} - \gamma_k^{(i)}}{\delta^{(i+1)}} \right),
\end{equation}
and $\delta^{(i+1)} = \mu \delta^{(i)}$ denotes the temperature parameter with cooling factor $\mu \in (0,1)$. Although global optimality cannot be guaranteed, this hybrid PSO–SA framework converges to high-quality solutions in practice.

\subsection{Beamforming Vector Design via MRT}
Given the optimized MA positions obtained from the previous subproblem, the corresponding channel vector $\mathbf{H}$ can be readily computed. 
With fixed antenna positions, the beamforming optimization subproblem admits a closed-form solution. 
Specifically, based on the conclusion of \cite{MRT1}, MRT can be employed to maximize the received SNR, yielding
\begin{equation}\label{eq:w}
\mathbf{w} = \sqrt{P_{\max}} \frac{\mathbf{H}}{\|\mathbf{H}\|},
\end{equation}
where $\|\mathbf{H}\|$ denotes the Euclidean norm of the vector $\mathbf{H}$.

\subsection{Overall Algorithm}

For Problem~$\mathcal{P}2$, with the position of ${\rm UAV}_t$ fixed, the resulting MA positioning problem is non-convex due to the irregular feasible region and the highly nonlinear dependence of the received SNR on antenna locations. 
To efficiently obtain a high-quality solution, we employ a hybrid PSO-SA algorithm, which combines the global search capability of PSO with the probabilistic escape mechanism of SA. 
The detailed pseudo-code of the proposed PSO--SA algorithm is summarized in Algorithm~\ref{tab:PSO_SA}.

Building upon Problem~$\mathcal{P}2$, we further consider the volumetric coverage maximization problem over a discretized 3D voxel space. 
A voxel is regarded as \emph{covered} if the maximum achievable SNR at the corresponding UAV location exceeds a predefined threshold $\epsilon$. 
Accordingly, for each voxel $t$, the SNR maximization problem $\mathcal{P}2$ serves as an inner feasibility-checking subproblem.

\begin{algorithm}[H] 
    \caption{PSO-based Antenna Position Optimization with SA (PSO--SA)}  
    \begin{algorithmic}[1]  
        \State \textbf{Initialization:} 
        Given the UAV position $\mathbf{p}_t$, initialize the MA positions $\mathbf{q}_k^{(0)}$, particle number $K$, maximum iteration number $I_{\max}$, and initial velocities $\mathbf{v}_k^{(0)}=\mathbf{0}$.
        Initialize the local best SNR $\gamma_{k,\mathrm{lbest}}$ and global best SNR $\gamma_{\mathrm{gbest}}$.

        \For{$i = 1$ to $I_{\max}$}
            \For{$k = 1$ to $K$}
                \State Update particle velocity using Eq. \eqref{eq:v_k};
                \State Apply velocity constraints using Eq. \eqref{eq:v_cons};
                \State Update particle position using Eq. \eqref{eq:q_k};
                \State Compute beamforming vector $\mathbf{w}$ using Eq. \eqref{eq:w};
                \State Evaluate the received SNR $\gamma_k^{(i)}$;
                \If{$\gamma_k^{(i)} > \gamma_{k,\mathrm{lbest}}$}
                    \State Update $\gamma_{k,\mathrm{lbest}} \leftarrow \gamma_k^{(i)}$;
                    \State Update $\mathbf{q}_{k,\mathrm{lbest}} \leftarrow \mathbf{q}_k^{(i)}$;
                \EndIf
            \EndFor

            \State Compute SA acceptance probability using Eq. \eqref{eq:p_SA};
            \If{$\gamma_{\mathrm{gbest}}$ is improved or $rand() < p_{\mathrm{SA}}$}
                \State Update global best solution $\mathbf{q}_{\mathrm{gbest}}$;
            \EndIf
        \EndFor
        \State \textbf{Output:} Global best antenna positions $\mathbf{q}_{\mathrm{gbest}}$ and corresponding SNR $\gamma_{\mathrm{gbest}}$.
    \end{algorithmic}
    \label{tab:PSO_SA}
\end{algorithm}

Specifically, for each voxel, the proposed PSO--SA algorithm is invoked to iteratively optimize the MA positions so as to maximize the received SNR. 
Once the resulting SNR satisfies the condition of $\gamma_t \ge \epsilon$, the voxel is declared covered and the optimization process is terminated early. 
This early-stopping mechanism significantly reduces the overall computational complexity without affecting the correctness of the coverage evaluation.
The complete procedure for solving Problem~$\mathcal{P}1$ is summarized in Algorithm~\ref{tab:cover}.

\begin{algorithm}[H] 
    \caption{Volumetric Coverage Maximization via PSO-SA}  
    \begin{algorithmic}[1]  
        \State \textbf{Initialization:} 
        Set the number of voxels $T$ and initialize coverage indicator $c_t = 0$, $\forall t$.

        \For{$t = 1$ to $T$}
            \State Apply Algorithm~\ref{tab:PSO_SA} to solve Problem~$\mathcal{P}2$ for voxel $t$;
            \If{the obtained SNR satisfies $\gamma_t \ge \varepsilon$}
                \State Set $c_t = 1$;
            \EndIf
        \EndFor

        \State \textbf{Output:} Coverage vector $\mathbf{c} = [c_1,\ldots,c_T]$.
    \end{algorithmic}
    \label{tab:cover}
\end{algorithm}

\subsection{Computational Complexity Analysis}

For Algorithm~\ref{tab:PSO_SA}, each PSO iteration involves updating particle velocities and positions, computing the MRT beamforming vector, and evaluating the received SNR. 
These operations scale linearly with the number of antenna elements $M$. 
Since the particle swarm contains $K$ particles, the computational complexity per iteration is $O(KM)$. 
With a maximum of $I_{\max}$ iterations, the overall time complexity of Algorithm~\ref{tab:PSO_SA} is $O(I_{\max} K M)$. 
The SA-based acceptance step introduces only constant-time operations per iteration and thus does not affect the overall complexity order.

For Algorithm~\ref{tab:cover}, the outer loop traverses all $T$ voxels in the discretized coverage region. 
For each voxel, Algorithm~\ref{tab:PSO_SA} is executed once to determine its coverability. 
Therefore, the total computational complexity of the proposed volumetric coverage maximization framework is $O(T I_{\max} K M)$.

\begin{table}[!htb]  
\caption{\small Simulated parameters and values}
\centering
\label{tab:sim para validation}
\begin{tabular}{p{0.7cm}p{5cm}p{2cm}}
\toprule
  Symbols & Meanings & Values \\
 
\midrule
   $M$          & number of antennas  & 9 ($3 \times 3$  array) \\
  $\mathbf{q}_{\rm BS}$          & position of base station     & $(0,0,10)$ \\
   $d_{\rm mov}$          &antenna movement range & $\{0.5\lambda, 5\lambda\}$\\
  $F$          &frequency band of communication & 3.5GHz \cite{Yujia_TCOM}\\
  $\lambda $   &  beamwave   & $0.0857$m \\
  $\alpha$        &loss path   & $2$ \\
  $P_{\rm max}$        & the maximum transmission power   & $40$dBm \\
  $L_{\rm NLoS}$     & number of NLoS   & $5$ \\
  $\epsilon$     & threshold of SNR    & $10$dB \\
  $P_{\max}$     & maximum communication power      & 40dBm \\
   $\kappa$     & Rician factor  & 3 \\
   $\beta$     & range of mechanical downtilt  & $[-20^\circ,20^\circ]$ \\
  $\rho$          & the radiation pattern sharpness factor  & $2$ \cite{G}  \\
  $\delta_0$   & initial temperature  & 1 \\
  $\mu$   &cooling factor & 0.98   \\ 
  \bottomrule
\end{tabular}
\end{table}

\section{Evaluations}\label{sec:sim}
This section conducts several numerical simulations to evaluate the communication performance for FPA and MA arrays within a defined region, where $x \in [0,1000]$, $y \in [-500,500]$, and $z \in [0,H]$, with $H \in \{ 300,600\}$.

To further validate the advantage of MA-enabled volumetric coverage, three benchmark schemes are considered for comparison: 1) \textbf{FPA\_noBF}: 
1) \textbf{FPA\_noBF}: A FPA array with a given mechanical downtilt, where no beamforming is applied; 
2) \textbf{FPA\_BF}: A FPA array with a given mechanical downtilt, where transmit beamforming is employed;  
3) \textbf{MA\_BF}: A MA array with a given UPA downtilt, where the 3D antenna positions and transmit beamforming are jointly optimized.
We first evaluate the received SNR performance of the above schemes for a given single-UAV location. 
Then, the coverability of the three schemes is compared with that of the \textbf{4DMA\_BF} scheme, where the 3D MA positions, the array mechanical downtilt and the transmit beamforming are jointly optimized.
The key simulation parameters are listed in \ref{tab:sim para validation}.

\begin{figure}
\centering
\includegraphics[width=2.8in]{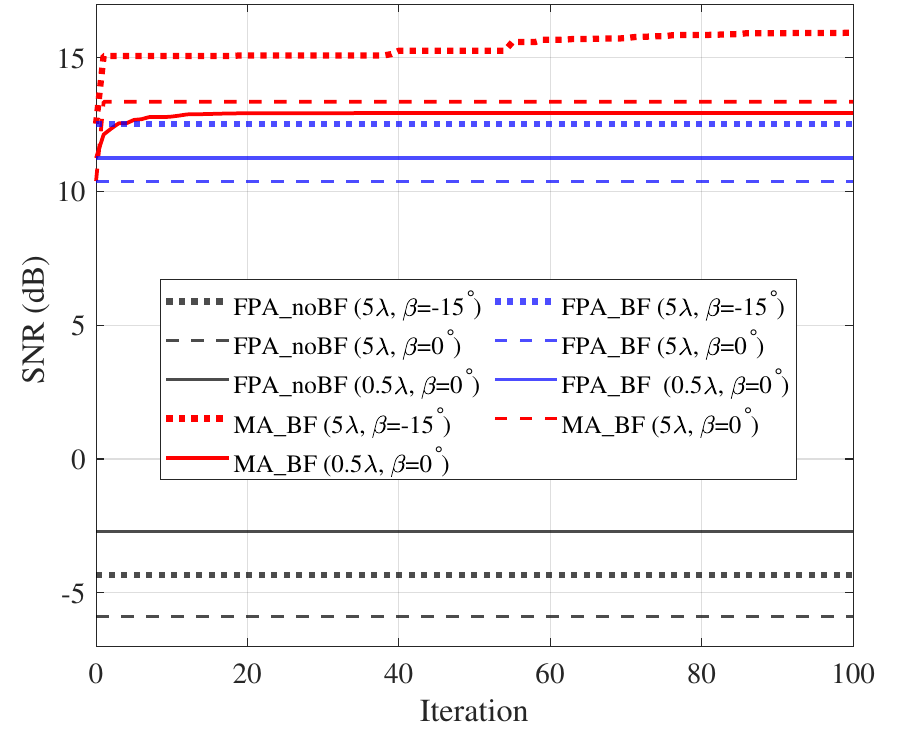}
\caption{\small iterations vs. SNR}
\label{fig:iter_SNR}
\end{figure}

\begin{figure}[!ht]
    \centering
    \begin{subfigure}{\linewidth}
        \centering
        \includegraphics[width=0.8\linewidth]{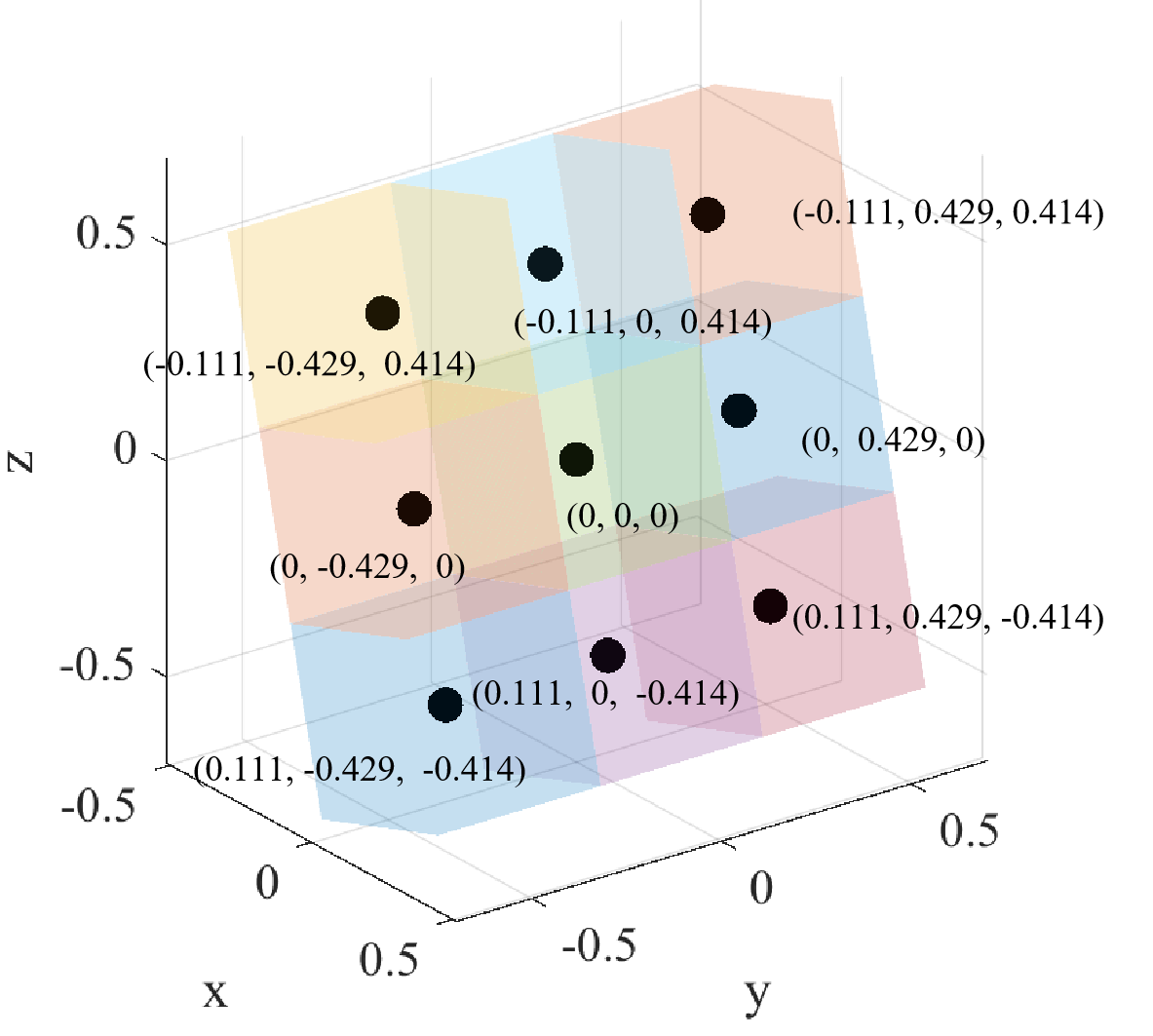}
        \caption{\small 3D positions for FPA system }
        \label{fig:FPA_positon}
    \end{subfigure}
    \\[1ex] 
    \begin{subfigure}{\linewidth}
        \hspace{1cm}
        \includegraphics[width=0.85\linewidth]{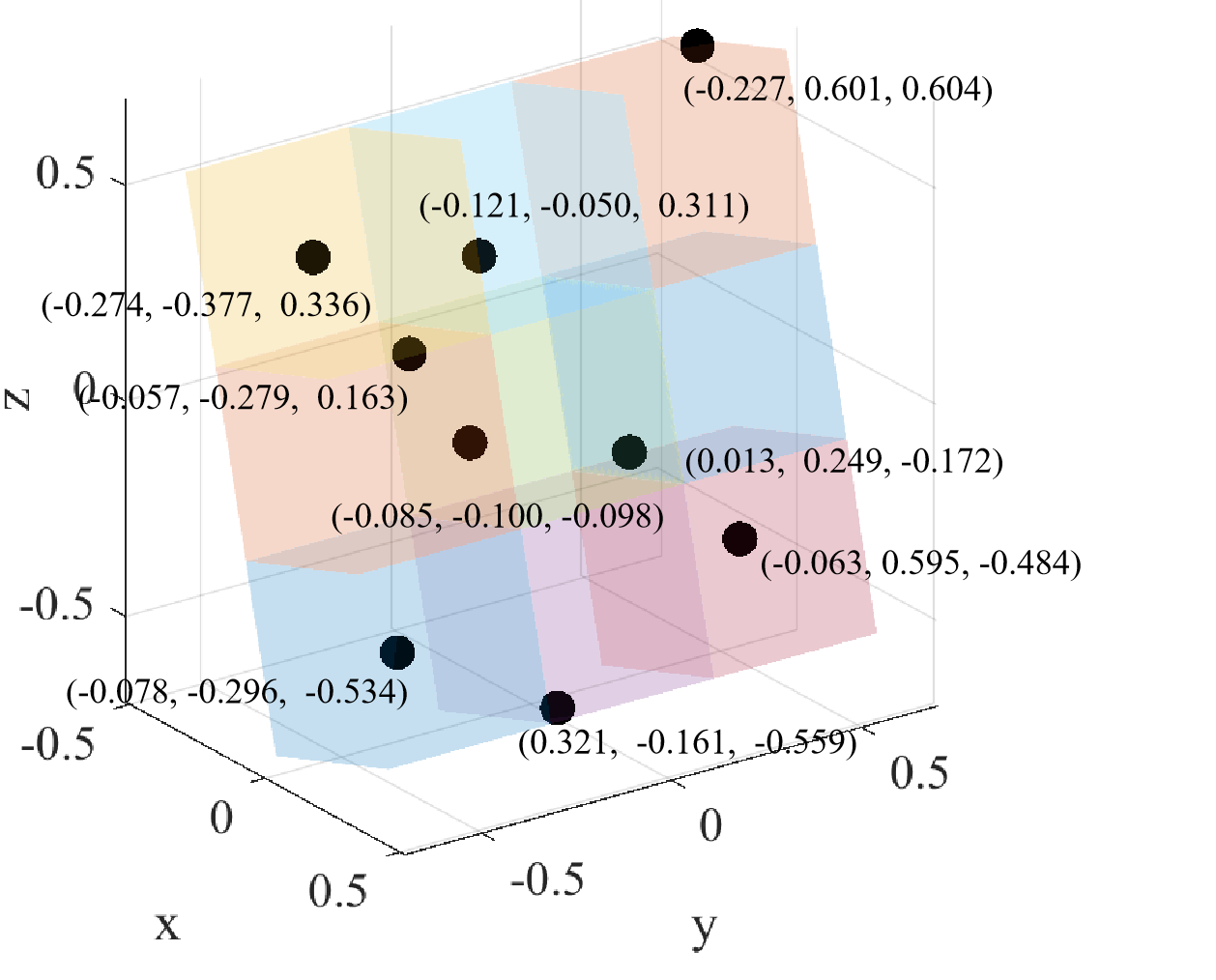}
        \caption{\small 3D positions for MA system}
        \label{fig:MA_positions}
    \end{subfigure}
    \caption{\small Optimized antenna positions with $\beta=-15^\circ$}
    \label{fig:position_15}
\end{figure}

\begin{figure}
\centering
\includegraphics[width=2.8in]{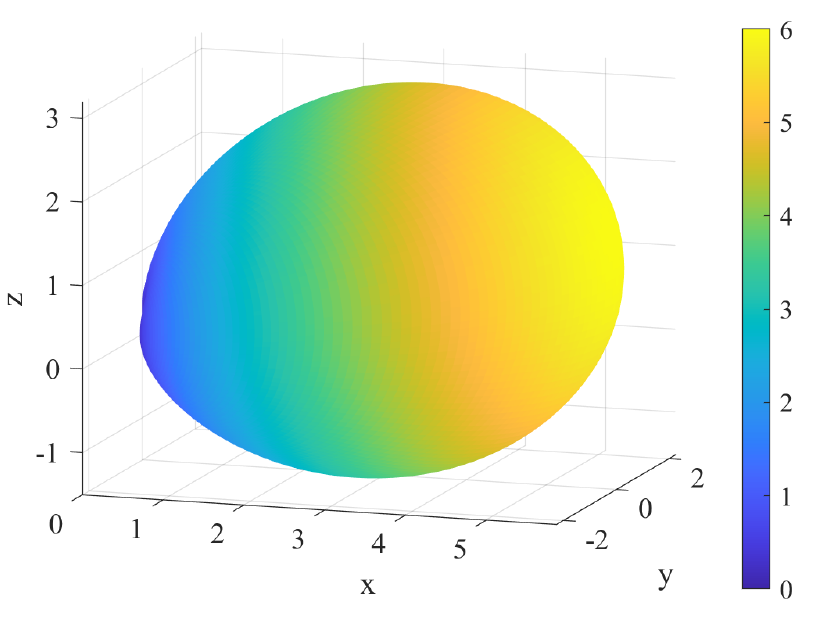}
\caption{\small 3D radiation pattern}
\label{fig:3D_Radiation_Pattern}
\end{figure}

\subsection{SNR Evaluation for a Given ${\rm UAV}_t$}
For a given location of ${\rm UAV}_t$ at $(400,\,200,\,300)$, the performance and convergence under different parameter settings are evaluated, as shown in Fig.\ref{fig:iter_SNR}. 
The proposed algorithm exhibits noticeable fluctuations in the initial iterations, but gradually converges as the number of iterations increases. 
It is evident that the \textbf{MA\_BF} scheme consistently achieves the highest SNR performance owing to its enhancement of spatial degrees of freedom. 
The \textbf{FPA\_BF} scheme can also attain satisfactory performance through beamforming design, whereas the \textbf{FPA\_noBF} scheme suffers from inferior performance due to its fixed mechanical structure and lack of beamforming flexibility.
For the FPA-based schemes, increasing the inter-element spacing from $0.5\lambda$ to $5\lambda$ can degrade the SNR due to the fixed array geometry. In contrast, for the MA-based schemes, allowing the antenna elements to move within a $5\lambda$ region provides significantly enhanced spatial degrees of freedom, which enables more flexible channel adaptation and SNR improvement.
Moreover, when the antenna mechanical downtilt is set to $-15^\circ$, corresponding to an upward-tilted array configuration, a noticeable SNR enhancement over the three-dimensional airspace is observed.
This observation indicates that the antenna mechanical downtilt should be flexibly designed according to the specific coverage region.

Fig. \ref{fig:position_15} illustrates the comparison between the antenna configurations for FPA and MA 3D positions when the downtilt angle is set to $-15^\circ$. 
With $\rho = 2$, the radiation patterns corresponding to downtilt angles of $-15^\circ$ is depicted in Fig.\ref{fig:3D_Radiation_Pattern} according to Eq. \eqref{eq:G}. 
It can be observed that the main radiation direction is perpendicular to the UPA plane.
When the mechanical downtilt is adjusted to $-15^\circ$, the antenna array is tilted upward, resulting in a corresponding shift of the main radiation direction.

\begin{figure*}[!ht]
    \centering
    \begin{subfigure}{\textwidth}
        \centering
        \includegraphics[width=0.8\linewidth]{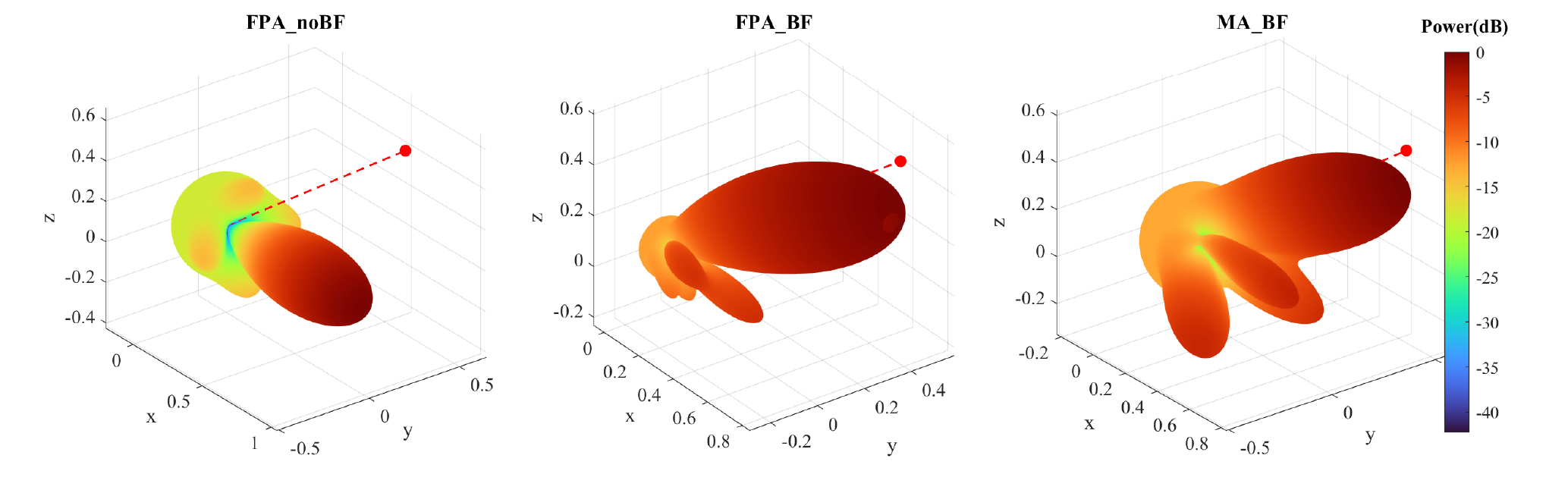}
        \caption{\small Antenna movement range: $0.5\lambda$}
        \label{fig:power_05}
    \end{subfigure}

    \begin{subfigure}{\textwidth}
        \centering
        \includegraphics[width=0.85\linewidth]{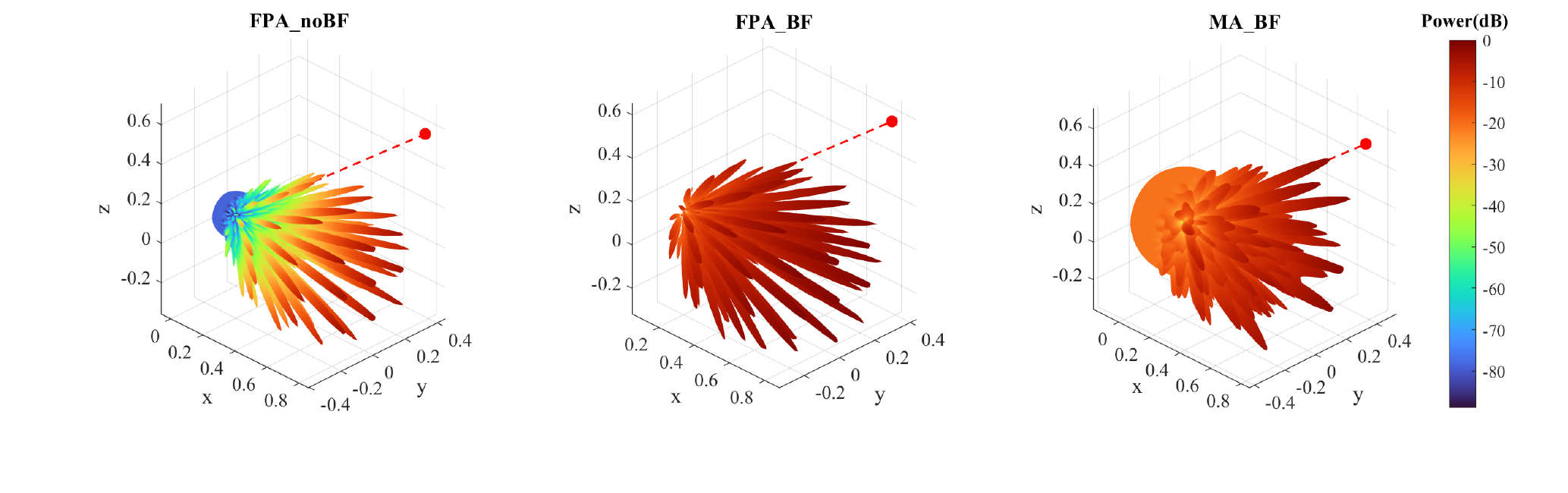}
        \caption{\small Antenna movement range: $5\lambda$}
        \label{fig:power_5}
    \end{subfigure}
    \caption{\small Power radiation pattern}
    \label{fig:power}
\end{figure*}

The power radiation pattern corresponding to $\beta =0^\circ$ is shown in Fig. \ref{fig:power}. 
For the \textbf{FPA\_BF} and \textbf{MA\_BF} schemes, the radiate power is effectively concentrated toward the receiver direction through flexible beamforming design. 
When the inter-element spacing or antenna movement range is set to $0.5\lambda$, the resulting beam becomes relatively wide, with a pronounced main lobe.
When the antenna movement range is expanded to $5\lambda$, the increased inter-element spacing leads to a significantly narrower main beam accompanied by more pronounced side lobes. Although both the \textbf{FPA\_BF} and \textbf{MA\_BF} schemes steer the beam toward the receiver direction, the \textbf{MA\_BF} scheme exhibits a more distinct antenna directivity due to the additional antenna position optimization.

\begin{figure*}[!ht]
    \centering

    \begin{subfigure}{\textwidth}
        \centering
        \begin{subfigure}{0.48\textwidth}
            \centering
            \includegraphics[width=\linewidth]{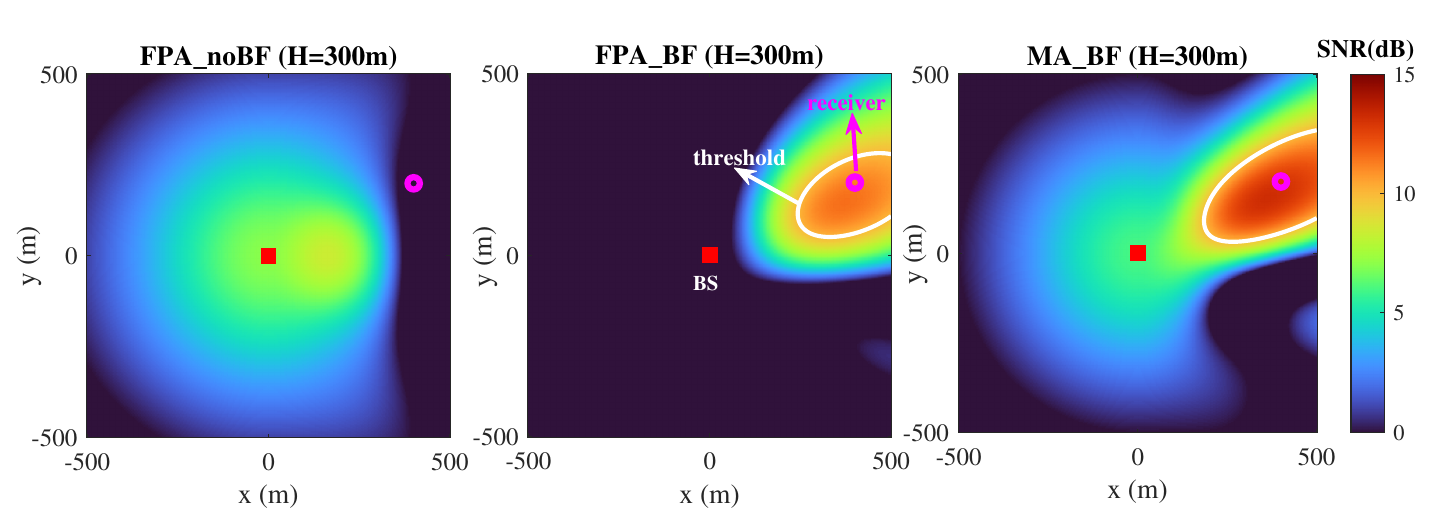}
        \end{subfigure}
        \hfill
        \begin{subfigure}{0.48\textwidth}
            \centering
            \includegraphics[width=\linewidth]{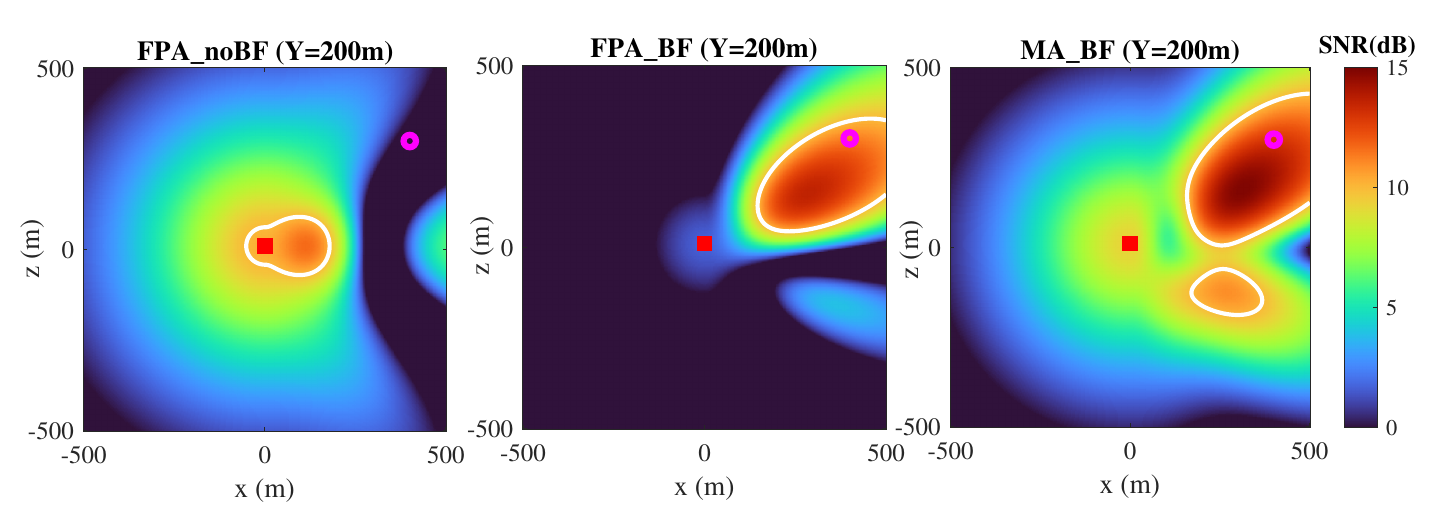}
        \end{subfigure}
        \caption{\small Antenna movement range: $0.5\lambda$, $\beta = 0^\circ$}
    \end{subfigure}

    \begin{subfigure}{\textwidth}
        \centering
        \begin{subfigure}{0.48\textwidth}
            \centering
            \includegraphics[width=\linewidth]{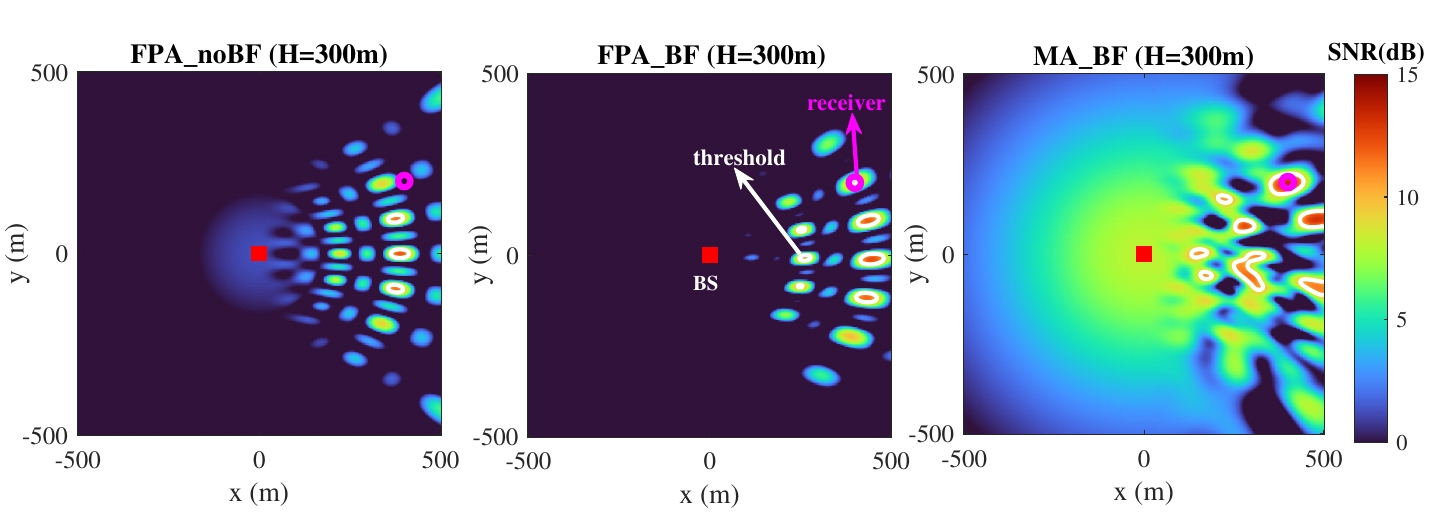}
        \end{subfigure}
        \hfill
        \begin{subfigure}{0.48\textwidth}
            \centering
            \includegraphics[width=\linewidth]{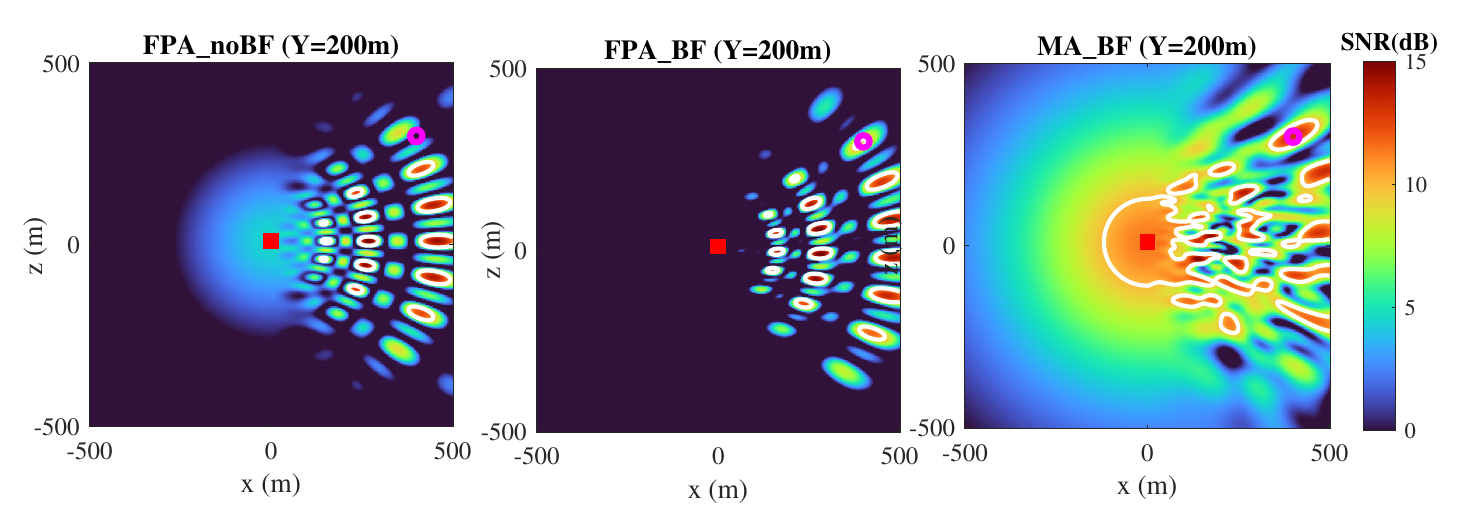}
        \end{subfigure}
        \caption{\small Antenna movement range: $5\lambda$, $\beta = 0^\circ$}
    \end{subfigure}

    \begin{subfigure}{\textwidth}
        \centering
        \begin{subfigure}{0.48\textwidth}
            \centering
            \includegraphics[width=\linewidth]{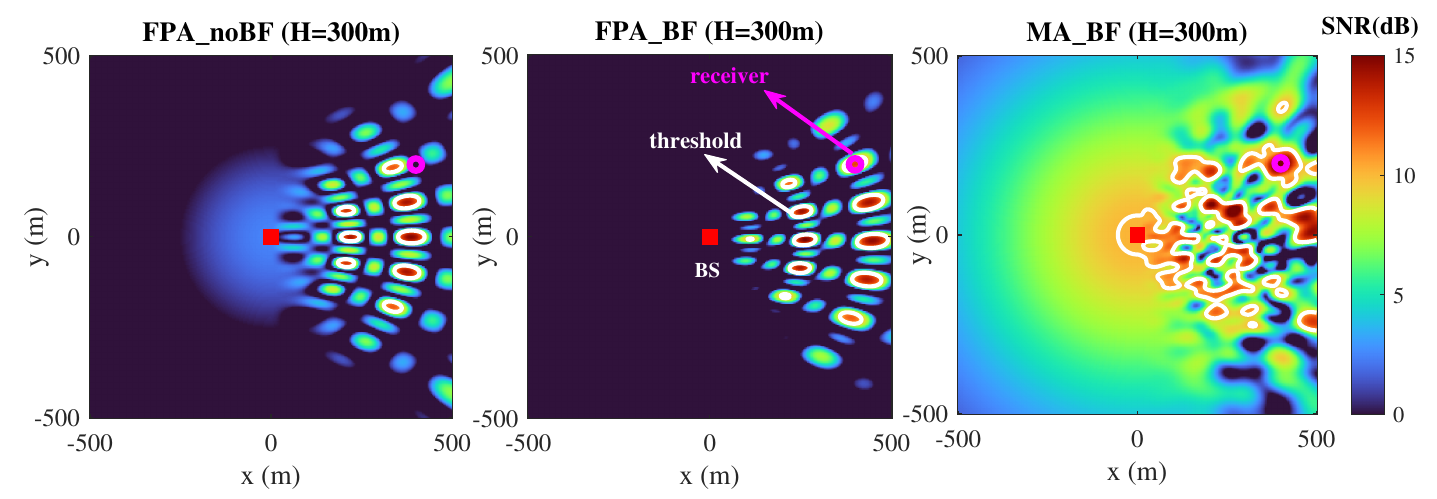}
        \end{subfigure}
        \hfill
        \begin{subfigure}{0.48\textwidth}
            \centering
            \includegraphics[width=\linewidth]{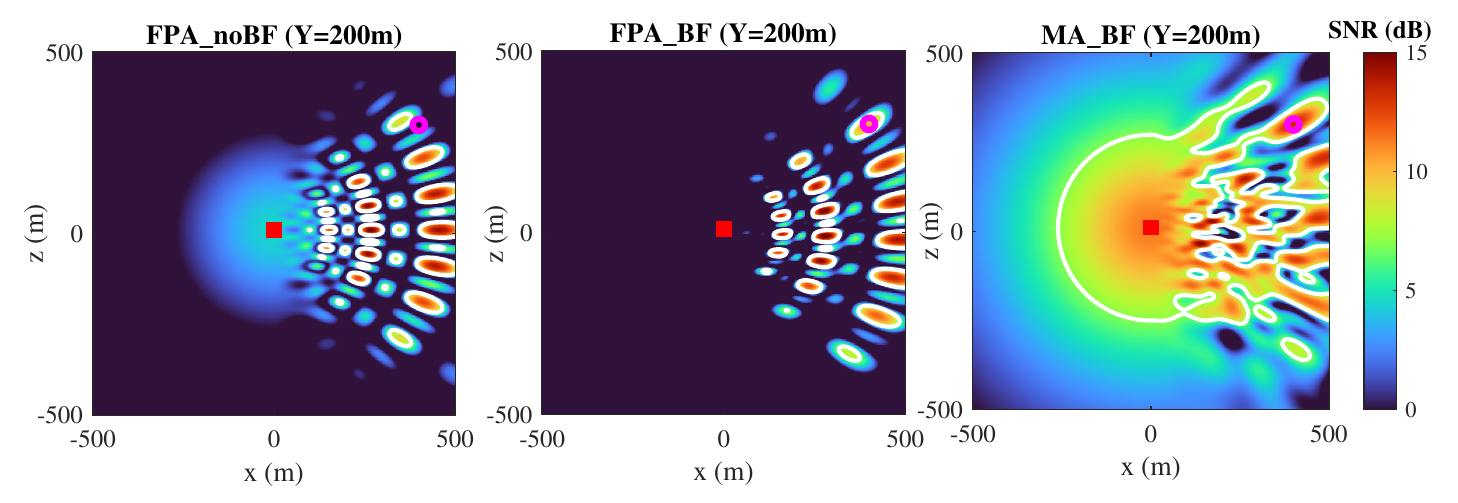}
        \end{subfigure}
        \caption{\small Antenna movement range: $5\lambda$, $\beta = -15^\circ$} \label{fig:snr_slices3}
    \end{subfigure}

    \caption{\small SNR distribution on $H=300$ and $Y=200$ slices}
    \label{fig:snr_slices}
\end{figure*}

The SNR distributions over different spatial directions for antenna movement ranges of $0.5\lambda$ and $5\lambda$ are illustrated in Fig. \ref{fig:snr_slices}, where cross-sectional slices along the planes $H = 300$m and $Y = 200$m are considered. 
It can be clearly observed that the scheme without beamforming yields SNR values below $0$dB at the receiver, indicating insufficient signal focusing capability. A larger movement range enables the antenna array to adjust its beam more flexibly, thereby dynamically adapting to varying signal demands and environmental conditions. As shown in Fig. \ref{fig:snr_slices3},  when the array is tilted upward ($\beta =-15^\circ$), the beam steering effect can be clearly observed from the $Y=200$ spatial slice, where a larger coverage region is achieved at the altitude of $H=300$m.

\subsection{3D Coverability Under Different Parameter Settings}

Firstly, we evaluate the 3D coverage of the three schemes with a fixed mechanical downtilt under different parameter settings. The coverage performance is examined over the region defined by $x \in [0,1000]$ and $y \in [-500,500]$, while considering the low-altitude airspace below $300~\mathrm{m}$ and $600~\mathrm{m}$, respectively.
Under the configuration with $d_{\rm mov}=5\lambda$ and $\beta = -15^\circ$, the resulting coverable regions are presented in Fig.~X and Fig.~X.
For the \textbf{FPA\_noBF} scheme, the coverage region is severely limited due to its fixed antenna positions and the absence of beamforming flexibility. 
In contrast, the \textbf{FPA\_BF} and \textbf{MA\_BF} schemes significantly expand the coverable region by exploiting adaptive transmit beamforming. 
By further leveraging the 3D mobility of MA elements, the \textbf{MA\_BF} scheme achieves additional degrees of freedom, attaining coverability ratios of $85.75\%$ and $72.69\%$ within altitude ranges of $0-300\,\mathrm{m}$ and $0-600\,\mathrm{m}$, respectively. 
Compared with the \textbf{FPA\_BF} scheme, the coverability ratios are improved by $26.8$ and $29.65$ percentage points within the $300\,\mathrm{m}$ and $600\,\mathrm{m}$ altitude ranges, respectively.

\begin{figure*}[!ht]
    \centering
    \begin{subfigure}{\textwidth}
        \centering
        \includegraphics[width=0.9\linewidth]{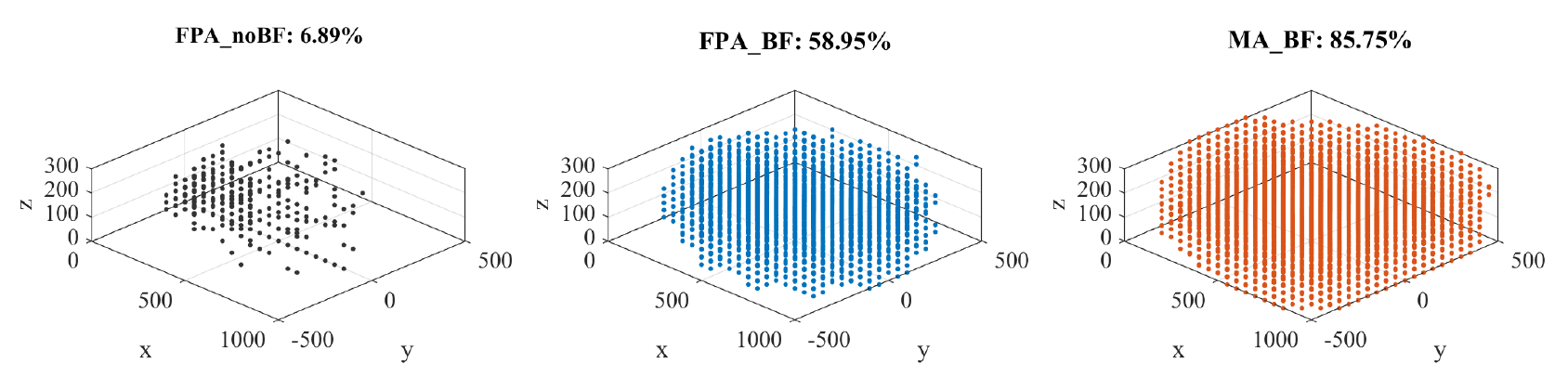}
        \caption{\small Coverage of the airspace below $300\,\mathrm{m}$
}
        \label{fig:Cover300}
    \end{subfigure}

    \begin{subfigure}{\textwidth}
        \centering
        \includegraphics[width=0.9\linewidth]{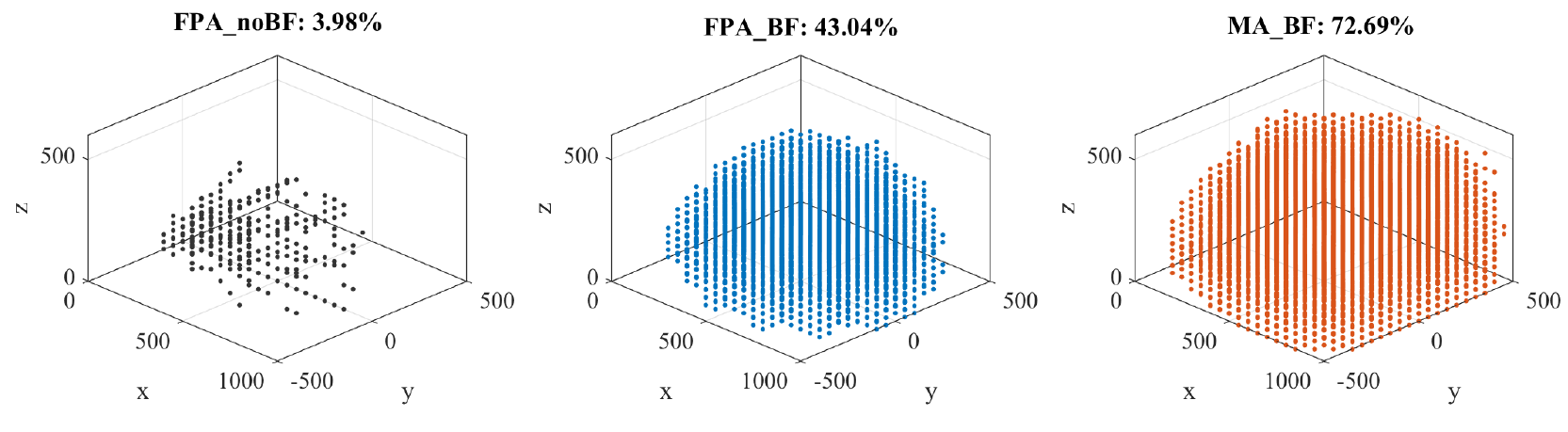}
        \caption{\small Coverage of the airspace below $600\,\mathrm{m}$}
        \label{fig:Cover600}
    \end{subfigure}

    \caption{\small 3D coverage under $d_{\rm mov}=5\lambda, \beta=-15^\circ$}
    \label{fig:Coverage}
\end{figure*}

\begin{figure*}[t]
    \centering
    \begin{subfigure}{0.45\linewidth}
        \centering
        \includegraphics[width=\linewidth]{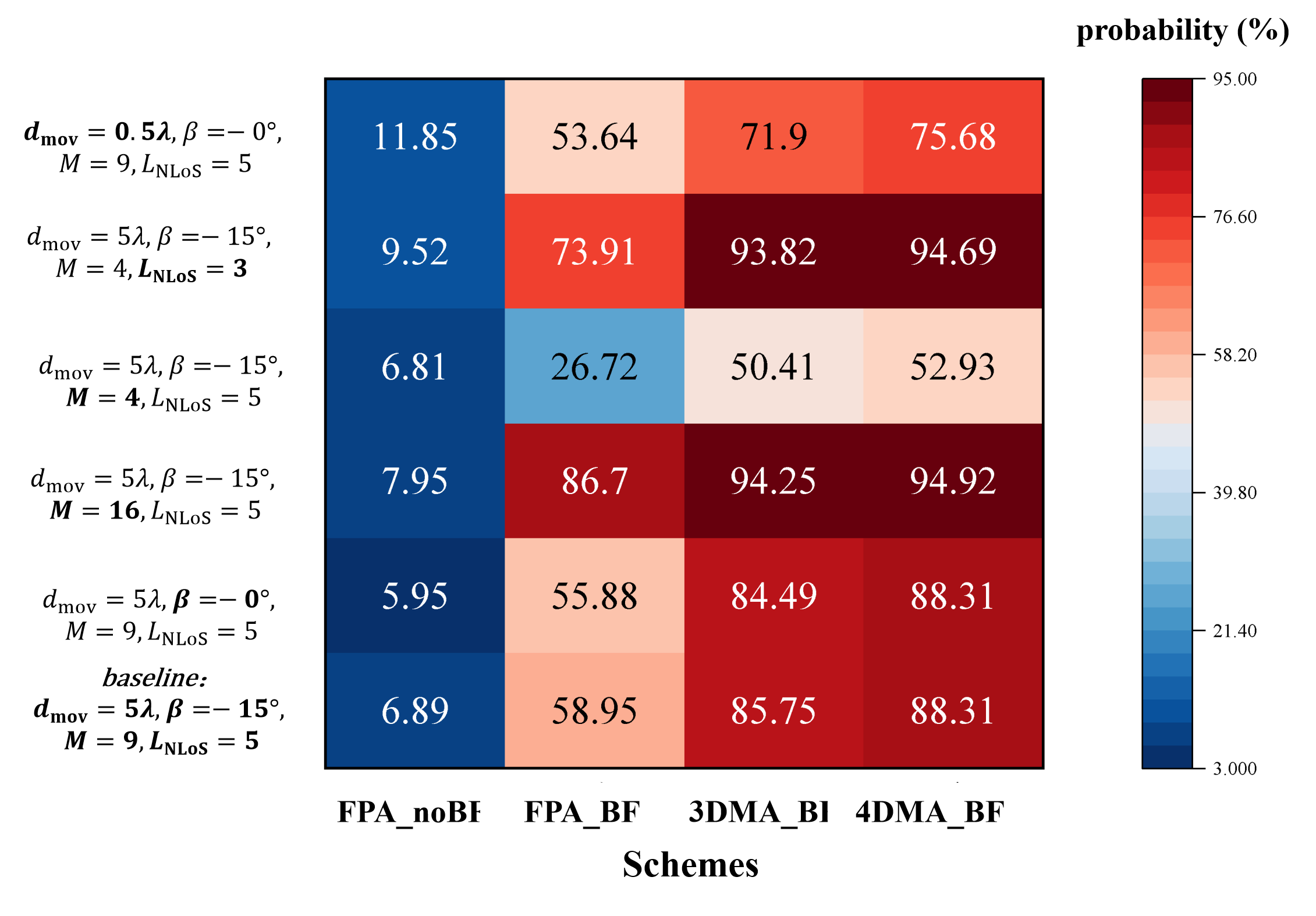}
        \caption{\small Coverage of the airspace below $300\,\mathrm{m}$}
        \label{fig:4DMA_300}
    \end{subfigure}
    \hspace{0.05\linewidth} 
    \begin{subfigure}{0.45\linewidth}
        \centering
        \includegraphics[width=\linewidth]{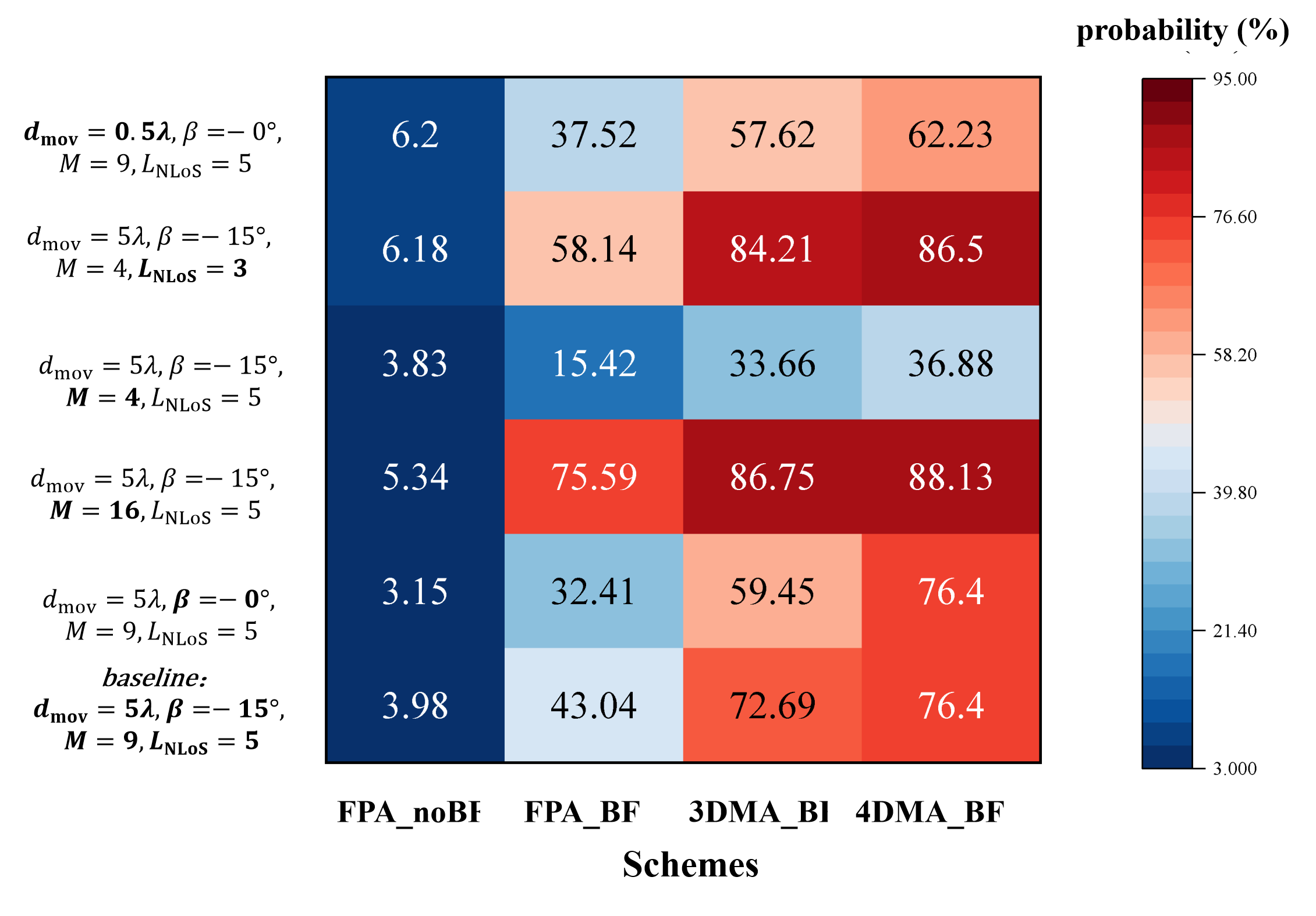}
        \caption{\small Coverage of the airspace below $600\,\mathrm{m}$}
        \label{fig:4DMA_600}
    \end{subfigure}
    \caption{\small Performance comparison of four schemes under different parameters}
    \label{fig:4DMA}
\end{figure*}

Furthermore, the \textbf{4DMA\_BF} control scheme is developed by extending the \textbf{MA\_BF} design to additionally optimize the mechanical downtilt, which is allowed to vary freely within the range $[-20^\circ,\,20^\circ]$. 
Under the aforementioned parameter settings, the \textbf{4DMA\_BF} scheme achieves 3D coverability ratios of $88.31\%$ and $76.4\%$ within the regions below altitudes of $300\,\mathrm{m}$ and $600\,\mathrm{m}$, respectively. 
Moreover, Fig. \ref{fig:4DMA} provides a detailed performance comparison of the four schemes under different parameter settings and leads to the following key observation. 

By comparison, significant conclusions can be drawn: 

1) The coverage ranking of the four schemes is given by
$\textbf{4DMA\_BF} > \textbf{3DMA\_BF} > \textbf{FPA\_BF} > \textbf{FPA\_noBF},$
which clearly demonstrates the superiority of MA-based schemes in enhancing volumetric coverability. The  \textbf{4DMA\_BF} design further improves the communication performance by exploiting additional spatial DoF.

2) Under a fixed mechanical downtilt constraint, an upward-tilted array configuration ($\beta = -15^\circ$) can effectively enhance the  coverability of the airspace.

3) Increasing the number of antenna elements consistently improves the coverage performance, which can be attributed to enhanced array gain, improved spatial resolution, and stronger spatial diversity, all of which are critical for spatial coverage enhancement.

4) Under the Rician channel model with $\kappa=3$, where the LoS component is dominant, an increase in the number of NLoS paths leads to a degradation in coverage performance. This is because the additional NLoS components introduce power dispersion and phase randomness, which dilute the effective LoS energy concentration within the target low-altitude airspace.

5) Enlarging the inter-element spacing further improves the 3D coverability below $300$m and $600$m, as it reduces spatial correlation among antenna elements and enables higher angular resolution, thereby facilitating more effective spatial energy reconstruction over the coverage volume.

\section{Conclusion}\label{sec:conclusion}
In this paper, we demonstrated the advantages of MAs in compensating for airspace coverage gaps. We formulated a spatial coverage maximization problem and developed efficient PSO-SA algorithms to jointly optimize the 3D MA positions and beamforming matrices. Extensive simulations quantified the coverage performance differences between MA- and FPA-based schemes under various system parameters. The results showed that, for a given mechanical tilt configuration, the proposed 3DMA design significantly improved the achievable coverage ratio. Notably, allowing flexible adjustment of the mechanical tilt further enhanced the coverage performance. Future work will investigate intelligent and closed-loop MA control strategies for dynamic airspace environments.

\bibliographystyle{IEEEtran}
\bibliography{ref}

\begin{thebibliography}{10}
\providecommand{\url}[1]{#1}
\csname url@samestyle\endcsname
\providecommand{\newblock}{\relax}
\providecommand{\bibinfo}[2]{#2}
\providecommand{\BIBentrySTDinterwordspacing}{\spaceskip=0pt\relax}
\providecommand{\BIBentryALTinterwordstretchfactor}{4}
\providecommand{\BIBentryALTinterwordspacing}{\spaceskip=\fontdimen2\font plus
\BIBentryALTinterwordstretchfactor\fontdimen3\font minus
  \fontdimen4\font\relax}
\providecommand{\BIBforeignlanguage}[2]{{%
\expandafter\ifx\csname l@#1\endcsname\relax
\typeout{** WARNING: IEEEtran.bst: No hyphenation pattern has been}%
\typeout{** loaded for the language `#1'. Using the pattern for}%
\typeout{** the default language instead.}%
\else
\language=\csname l@#1\endcsname
\fi
#2}}
\providecommand{\BIBdecl}{\relax}
\BIBdecl

\bibitem{coverage2}
Y.~Wang, H.~Zhao, H.~Huang, D.~Li, Y.~Ni, and G.~Gui, ``Multi-task multi-agent
  reinforcement learning for collaborative radio mapping and navigation in
  cellular-connected uav networks,'' \emph{IEEE Transactions on Cognitive
  Communications and Networking}, vol.~12, pp. 4731--4745, 2026.

\bibitem{coverage1}
Z.~Sheng, X.~Bowen, S.~Daohong, F.~Wei, J.~Zhiyuan, and N.~Zhisheng, ``Agile
  coverage for low-altitude aerial intelligent networks: A blended
  hyper-cellular solution,'' \emph{China Communications}, vol.~22, no.~9, pp.
  22--36, 2025.

\bibitem{range1}
X.~University, C.~M.~R. Institute, N.~U. of~Aeronautics, Astronautics, and
  P.~C.~L. and, ``White paper on digital low - altitude network architecture ,
  coverage and key technologies,'' \emph{Global 6G Conference}, 2025.

\bibitem{Wenxu1}
\BIBentryALTinterwordspacing
K.~Yu, W.~Wang, X.~Liu, Y.~Zhao, Q.~Zhang, Z.~Feng, and D.~Li, ``Does movable
  antenna present a dual-edged nature? from the perspective of physical layer
  security: A joint design of fixed-position antenna and movable antenna,''
  2025. [Online]. Available: \url{https://arxiv.org/abs/2507.05784}
\BIBentrySTDinterwordspacing

\bibitem{Yu_Survey}
K.~Yu, K.~Li, Y.~Zhao, Z.~Feng, D.~Li, Q.~Zhang, and J.~Yu, ``Movable
  antenna-aided secure v2x communication: An integrated sensing and
  communication perspective,'' \emph{IEEE Wireless Communications}, vol.~32,
  no.~6, pp. 118--124, 2025.

\bibitem{Shao20256D}
X.~Shao, Q.~Jiang, and R.~Zhang, ``6d movable antenna based on user
  distribution: Modeling and optimization,'' \emph{IEEE Transactions on
  Wireless Communications}, vol.~24, no.~1, pp. 355--370, 2025.

\bibitem{Cellular2019}
X.~Xu and Y.~Zeng, ``Cellular-connected uav: Performance analysis with 3d
  antenna modelling,'' in \emph{2019 IEEE International Conference on
  Communications Workshops (ICC Workshops)}, 2019, pp. 1--6.

\bibitem{Network2019}
J.~Lyu and R.~Zhang, ``Network-connected uav: 3-d system modeling and coverage
  performance analysis,'' \emph{IEEE Internet of Things Journal}, vol.~6,
  no.~4, pp. 7048--7060, 2019.

\bibitem{Maeng}
S.~J. Maeng, M.~M.~U. Chowdhury, s.~Güvenç, A.~Bhuyan, and H.~Dai, ``Base
  station antenna uptilt optimization for cellular-connected drone corridors,''
  \emph{IEEE Transactions on Aerospace and Electronic Systems}, vol.~59, no.~4,
  pp. 4729--4737, 2023.

\bibitem{Research2025}
J.~Yu, W.~Xie, X.~Ding, J.~Li, and Q.~Bi, ``Research and improvement methods on
  capacity of air-ground integrated network based on vertical stratification,''
  \emph{Science China Information Sciences}, pp. 1--22, 2025.

\bibitem{Telecommunication2022}
B.~Li, Y.~Hu, X.~Wang, G.~Xu, G.~Liu, T.~Xiao, C.~Cheng, and Y.~Li, ``Research
  on application of 5g network ssb 1+x beamforming technology,''
  \emph{Telecommunication Sciences}, vol.~38, no.~1, pp. 150--158, 2022.

\bibitem{Wei1}
M.~Wei and W.~XIE, ``Research on low-altitude 3.5 ghz dual-carrier network
  coverage solution,'' \emph{Telecommunications Science}, vol.~41, no.~3, pp.
  87--95, 2025.

\bibitem{Wei2}
M.~Wei, Y.~Zhao, and J.~Yu, ``Research on low altitude network coverage
  solutions,'' in \emph{2025 International Wireless Communications and Mobile
  Computing (IWCMC)}, 2025, pp. 412--417.

\bibitem{Dedicating2023}
L.~Chen, M.~A. Kishk, and M.-S. Alouini, ``Dedicating cellular infrastructure
  for aerial users: Advantages and potential impact on ground users,''
  \emph{IEEE Transactions on Wireless Communications}, vol.~22, no.~4, pp.
  2523--2535, 2023.

\bibitem{ZTE2020}
Z.~Corporation, ``5g massive mimo network application,'' \emph{white paper},
  2020.

\bibitem{Technical2025}
S.~A.~V. ASSOCIATION, ``Technical specification for 5g network planning and
  construction to support low-altitude intelligent network services,'' 2025.

\bibitem{Lin2025}
L.~Lin, Y.~Liu, L.~Cao, J.~Gao, and T.~Jiang, ``Low-altitude communication
  networks planning for air routes coverage based on semi-supervised gcn,''
  \emph{IEEE Communications Letters}, vol.~29, no.~9, pp. 2178--2182, 2025.

\bibitem{ChinaM2021}
C.~MOBILE and e.~a. ZTE~CORPORATION, ``White paper on 3d coverage network for
  uav based on 5g communication technology,'' \emph{White paper}, 2021.

\bibitem{China2024}
C.~T.~R. INSTITUTE, ``White paper on air-ground integrated 5g enhanced
  low-altitude network design,'' \emph{White paper}, 2024.

\bibitem{Flexible2024}
D.~Wang, W.~Mei, B.~Ning, and Z.~Chen, ``Flexible beam coverage optimization
  for movable-antenna array,'' in \emph{IEEE Global Communications Conference},
  2024, pp. 4914--4919.

\bibitem{Gao_2026}
\BIBentryALTinterwordspacing
Y.~Gao, Q.~Wu, W.~Mei, G.~Chen, W.~Chen, and Z.~Zheng, ``Integrating movable
  antennas and intelligent reflecting surfaces for coverage enhancement,''
  \emph{IEEE Transactions on Wireless Communications}, vol.~25, p. 6082–6095,
  2026. [Online]. Available: \url{http://dx.doi.org/10.1109/TWC.2025.3623474}
\BIBentrySTDinterwordspacing

\bibitem{6DRen}
T.~Ren, X.~Zhang, L.~Zhu, W.~Ma, X.~Gao, and R.~Zhang, ``6-d movable antenna
  enhanced interference mitigation for cellular-connected uav communications,''
  \emph{IEEE Wireless Communications Letters}, vol.~14, no.~6, pp. 1618--1622,
  2025.

\bibitem{Yujia_TCOM}
Z.~Feng, Y.~Zhao, K.~Yu, and D.~Li, ``Movable antenna empowered pls with
  eve’s location uncertainty: Joint optimization of beamforming and antenna
  positions,'' \emph{IEEE Transactions on Communications}, vol.~73, no.~12, pp.
  13\,708--13\,724, 2025.

\bibitem{G}
S.~Yang, J.~An, Y.~Xiu, W.~Lyu, B.~Ning, Z.~Zhang, M.~Debbah, and C.~Yuen,
  ``Flexible antenna arrays for wireless communications: Modeling and
  performance evaluation,'' in \emph{2024 IEEE 24th International Conference on
  Communication Technology (ICCT)}, 2024, pp. 2034--2039.

\bibitem{GV2}
\BIBentryALTinterwordspacing
X.~Peng, Q.~Wu, Z.~Zheng, W.~Chen, Y.~Zhu, and Y.~Gao, ``Rotatable antenna
  enabled spectrum sharing: Joint antenna orientation and beamforming design,''
  2025. [Online]. Available: \url{https://arxiv.org/abs/2509.19912}
\BIBentrySTDinterwordspacing

\bibitem{GV1}
\BIBentryALTinterwordspacing
Y.~Zhang, Y.~Zhang, L.~Zhu, S.~Xiao, W.~Tang, Y.~C. Eldar, and R.~Zhang,
  ``6dma-aided hybrid beamforming with joint antenna position and orientation
  optimization,'' 2024. [Online]. Available:
  \url{https://arxiv.org/abs/2412.17088}
\BIBentrySTDinterwordspacing

\bibitem{Ricain1}
M.~Kang and M.~Alouini, ``Capacity of mimo rician channels,'' \emph{IEEE
  Transactions on Wireless Communications}, vol.~5, no.~1, pp. 112--122, 2006.

\bibitem{Ricain2}
P.~Liu, D.~Kong, J.~Ding, Y.~Zhang, K.~Wang, and J.~Choi, ``Channel estimation
  aware performance analysis for massive mimo with rician fading,'' \emph{IEEE
  Transactions on Communications}, vol.~69, no.~7, pp. 4373--4386, 2021.

\bibitem{Kaixuan_TMC}
K.~Li, K.~Yu, D.~Ma, Y.~Zhao, X.~Liu, Q.~Zhang, and Z.~Feng, ``Can movable
  antenna-enabled micro-mobility replace uav-enabled macro-mobility? a physical
  layer security perspective,'' \emph{IEEE Transactions on Mobile Computing},
  vol.~25, no.~3, pp. 4317--4330, 2026.

\bibitem{PSO1}
Y.~Qi and M.~Vaezi, ``Irs-assisted physical layer security in mimo-noma
  networks,'' \emph{IEEE Communications Letters}, vol.~27, no.~3, pp. 792--796,
  2023.

\bibitem{PSO2}
J.~Ding, Z.~Zhou, and B.~Jiao, ``Movable antenna-aided secure full-duplex
  multi-user communications,'' \emph{IEEE Transactions on Wireless
  Communications}, vol.~24, no.~3, pp. 2389--2403, 2025.

\bibitem{MRT1}
W.~Mei, X.~Wei, B.~Ning, Z.~Chen, and R.~Zhang, ``Movable-antenna position
  optimization: A graph-based approach,'' \emph{IEEE Wireless Communications
  Letters}, vol.~13, no.~7, pp. 1853--1857, 2024.

\end{thebibliography}
\end{document}